\documentclass[]{aastex63}

\newcommand\subs[1]{\textsubscript{#1}}
\newcommand\sups[1]{\textsuperscript{#1}}
\newcommand\rh[1]{\textcolor{black}{{\textit{r}\subs{H}}#1}}
\newcommand\lp[1]{\textcolor{black}{{\textit{L}\subs{p}}#1}}
\usepackage{ulem}

\received{April 6, 2021}
\revised{May 21, 2021}
\accepted{May 22, 2021}
\submitjournal{ApJ}

\shorttitle{}
\shortauthors{Roth et al.}
\begin{document}

%\title{Leveraging the ALMA Atacama Compact Array for Cometary Science: Mapping HCN, CH\subs{3}OH, H\subs{2}CO, CS, and HNC in Comet C/2015 ER61 (PanSTARRS)}

\title{Leveraging the ALMA Atacama Compact Array for Cometary Science: An Interferometric Survey of Comet C/2015 ER61 (PanSTARRS) and Evidence for a Distributed Source of Carbon Monosulfide}

\correspondingauthor{Nathan X. Roth}
\email{nathaniel.x.roth@nasa.gov}

\author[0000-0002-6006-9574]{Nathan X. Roth}
\affiliation{Solar System Exploration Division, Astrochemistry Laboratory Code 691, NASA Goddard Space Flight Center, 8800 Greenbelt Rd, Greenbelt, MD 20771, USA}
\affiliation{Department of Physics, The Catholic University of America, 620 Michigan Ave., N.E. Washington, DC 20064, USA}

\author [0000-0001-7694-4129]{Stefanie N. Milam}
\affiliation{Solar System Exploration Division, Astrochemistry Laboratory Code 691, NASA Goddard Space Flight Center, 8800 Greenbelt Rd, Greenbelt, MD 20771, USA}

\author[0000-0001-8233-2436]{Martin A. Cordiner}
\affiliation{Solar System Exploration Division, Astrochemistry Laboratory Code 691, NASA Goddard Space Flight Center, 8800 Greenbelt Rd, Greenbelt, MD 20771, USA}
\affiliation{Department of Physics, The Catholic University of America, 620 Michigan Ave., N.E. Washington, DC 20064, USA}

\author{Dominique Bockelée-Morvan}
\affiliation{LESIA, Observatoire de Paris, Université PSL, CNRS, Sorbonne Université,
Université de Paris, 5 place Jules Janssen, F-92195 Meudon, France}

\author[0000-0003-2414-5370]{Nicolas Biver}
\affiliation{LESIA, Observatoire de Paris, Université PSL, CNRS, Sorbonne Université,
Université de Paris, 5 place Jules Janssen, F-92195 Meudon, France}

\author[0000-0002-1545-2136]{Jérémie Boissier}
\affiliation{Institut de Radioastronomie Millimetrique, 300 rue de la pPiscine, F-38406
Saint Martin d'Heres, France}

\author[0000-0002-0500-4700]{Dariusz C. Lis}
\affiliation{Jet Propulsion Laboratory, California Institute of Technology, 4800 Oak Grove Drive, Pasadena, CA 91109, USA}

\author[0000-0001-9479-9287]{Anthony J. Remijan}
\affiliation{National Radio Astronomy Observatory, 520 Edgemont Rd, Charlottesville, VA 22903, USA}

\author[0000-0001-6752-5109]{Steven B. Charnley}
\affiliation{Solar System Exploration Division, Astrochemistry Laboratory Code 691, NASA Goddard Space Flight Center, 8800 Greenbelt Rd, Greenbelt, MD 20771, USA}



%\nocollaboration{0}

\begin{abstract}

We report the first survey of molecular emission from cometary volatiles using standalone Atacama Compact Array (ACA) observations from the Atacama Large Millimeter/Submillimeter Array (ALMA) toward comet C/2015 ER61 (PanSTARRS) carried out on UT 2017 April 11 and 15, shortly after its April 4 outburst. These measurements of HCN, CS, CH\subs{3}OH, H\subs{2}CO, and HNC (along with continuum emission from dust) probed the inner coma of C/2015 ER61, revealing asymmetric outgassing and discerning parent from daughter/distributed source species. This work presents spectrally integrated flux maps, autocorrelation spectra, production rates, and parent scale lengths for each molecule and a stringent upper limit for CO. HCN is consistent with direct nucleus release in C/2015 ER61, whereas CS, H\subs{2}CO, HNC, and potentially CH$_3$OH are associated with distributed sources in the coma.  Adopting a Haser model, parent scale lengths determined for H\subs{2}CO (\textit{L}\subs{p} $\sim$ 2200 km) and HNC (\textit{L}\subs{p} $\sim$ 3300 km) are consistent with previous work in comets, whereas significant extended source production (\textit{L}\subs{p} $\sim$ 2000 km) is indicated for CS, suggesting production from an unknown parent in the coma. The continuum presents a point-source distribution with a flux density implying an excessively large nucleus, inconsistent with other estimates of the nucleus size. It is best explained by the thermal emission of slowly moving outburst ejectas, with total mass 5--8 $\times$ 10$^{10}$ kg. These results demonstrate the power of the ACA for revealing the abundances, spatial distributions, and locations of molecular production for volatiles in moderately bright comets such as C/2015 ER61.

\end{abstract}

\keywords{Molecular spectroscopy (2095) ---
High resolution spectroscopy (2096) --- Radio astronomy (1338) --- Comae (271) --- Radio interferometry(1346) --- Comets(280)}


\section{Introduction} \label{sec:intro}
Comets are among the most primitive remnants of solar system formation. Having accreted during the early stages of the protosolar disk, their initial composition should be indicative of the chemistry and thermodynamical processes present in the cold disk midplane where they formed \citep{Bockelee2004}. Shortly after their formation, the migration of the giant planets gravitationally scattered comets across the solar system, placing many in their present day dynamical reservoirs, the Kuiper disk or the Oort cloud \citep{Gomes2005,Morbidelli2005,Levison2011}. Thus, comets have been stored in the cold outer solar system for the last $\sim$4.5 billion years in a relatively quiescent and unaltered state, preserved as ``fossils'' of solar system formation. Characterizing the composition of the comet nucleus, predominantly inferred through remote sensing of coma gases as comets pass through the inner solar system (heliocentric distances \textit{r}\subs{H} $<$ 3 au), affords the opportunity to extract clues to solar system formation imprinted in their volatile composition.

Comet observations are interpreted using a Haser model \citep{Haser1957} with coma volatiles identified as ``parent'' or ``primary'' (produced via direct sublimation from the nucleus and thus directly indicative of its composition), as ``product'' or ``daughter'' (produced by photolysis of gas-phase species in the coma), or as a distributed source \citep[these cannot be explained by gas-phase processes and are likely formed from photo and thermal degradation of dust;][]{Cottin2008}. Such observations at multiple wavelengths have revealed extensive chemical diversity among the comet population independent of dynamical reservoir \citep[e.g.,][]{AHearn1995,Crovisier2009,Cochran2012,Ootsubo2012,DelloRusso2016}. In particular, spectroscopy at millimeter/submillimeter (mm/sub-mm) wavelengths has detected a wealth of coma molecules through their rotational transitions, including those associated with direct nucleus sublimation in some comets (e.g., hydrogen cyanide, HCN, and methanol, CH$_3$OH), species associated with production from distributed coma sources (e.g., formaldehyde, H$_2$CO, and hydrogen isocyanide, HNC), and complex organic molecules \citep[e.g., glycoaldehyde, C$_2$H$_4$O$_2$, and ethyl alcohol, C$_2$H$_5$OH;][]{Bockelee1994,Biver1999,Milam2006,Fray2006,Boissier2007,Lis2008,Biver2015}. The advent of the Atacama Large Millimeter/Sub-millimeter Array (ALMA) has made mapping the 3D distribution of volatiles in the inner coma with exceptional spatial and spectral resolution possible, enabling unambiguous distinction between parent and product species \citep{Cordiner2014,Bogelund2017,Cordiner2017a,Cordiner2019,Cordiner2020}. Although the long baselines (up to 16 km) provided by the 50 $\times$ 12 m antennas of the main ALMA array provide sensitive measurements at high angular resolution, they may resolve out the extended flux of most species and are blind to large-scale structures in the coma.

In contrast, the 12 $\times$ 7 m antennas of the ALMA Atacama Compact Array (ACA) are sensitive to extended molecular flux in the coma, providing short baselines between 9 and $\sim$50 m (corresponding to an angular resolution range from 3$\farcs$6 to 19$\farcs$3 at 345 GHz, equivalent to $\sim$3100 km -- 16,700 km at the geocentric distance of C/2015 ER61 during our observations), that is available year-round as opposed to the vast configuration changes throughout the year for the 12 m array. \cite{Roth2021a} reported the first detection of molecular emission in cometary atmospheres using the ACA, demonstrating rapidly varying, anisotropic CH$_3$OH production in comet 46P/Wirtanen via analysis of the spectral line profiles. Here we report the first comprehensive study of coma volatiles with the ALMA ACA in observations toward comet C/2015 ER61 (PanSTARRS) taken on UT 2017 April 11 and 15, shortly after its UT 2017 April 4 outburst. Secure detections of molecular emission were identified from HCN, CH\subs{3}OH, H\subs{2}CO, carbon monosulfide (CS), and HNC as well as continuum emission from dust. A stringent (3$\sigma$) upper limit was determined for carbon monoxide (CO) production. We report production rates, mixing ratios (relative to H\subs{2}O), spatial maps, and autocorrelation spectra of detected species. Section~\ref{sec:obs} discusses the observations and data analysis. Section~\ref{sec:results} presents our results from these data. Section~\ref{sec:prod} discusses the distribution of each molecule and compares our results to comets characterized to date, and Section~\ref{sec:cont} analyzes continuum emission in C/2015 ER61 (PanSTARRS).

\begin{deluxetable*}{ccccccccccccc}
\tablenum{1}
\tablecaption{Observing Log\label{tab:obslog}}
\tablewidth{0pt}
\tablehead{
\colhead{Setting} & \colhead{UT Date} & \colhead{UT Time} & \colhead{\textit{t}\subs{int}} &
\colhead{\textit{r}\subs{H}} & \colhead{$\Delta$} & \colhead{$\phi$} & \colhead{$\nu$} &
\colhead{\textit{N}\subs{ants}}  & \colhead{Baselines} & \colhead{$\theta$\subs{INT}} & \colhead{$\theta$\subs{AC}} & \colhead{PWV} \\
\colhead{} & \colhead{} & \nocolhead{} & \colhead{(minutes)} & \colhead{(au)} &
\colhead{(au)} & \colhead{($\degr$)} & \colhead{(GHz)}  & \colhead{} & \colhead{(m)}  &
\colhead{($\arcsec$)} & \colhead{($\arcsec$)} & \colhead{(mm)}
}
\startdata
1 & 2017 Apr 11 & 15:15--15:52 & 25 & 1.14 & 1.19 & 51 & 349.2 & 9 & 8.9--47.9 & 6.07$\times$2.56 & 28.35 & 1.14 \\
2 & 2017 Apr 15 & 10:45--11:43 & 37 & 1.12 & 1.18 & 51 & 357.4 & 12 & 8.8--48.9 & 4.89$\times$2.31 & 27.71 & 2.39
\enddata
\tablecomments{\textit{T}\subs{int} is the total on-source integration time. \textit{r}\subs{H}, $\Delta$, and $\phi$ are the heliocentric distance, geocentric distance,
and phase angle (Sun-comet-Earth), respectively, of ER61 at the time of observations. $\nu$ is the mean frequency of each instrumental setting. \textit{N}\subs{ants}
is the number of antennas utilized during each observation, with the range of baseline lengths indicated for each. $\theta$\subs{INT} is the angular resolution (Gaussian
fit to the PSF) for the synthesized beam at $\nu$, $\theta$\subs{AC} is the angular resolution of the autocorrelation (7 m primary) beam at $\nu$, and PWV is the mean precipitable water vapor at zenith during the observations.}
\end{deluxetable*}

\section{Observations and Data Reduction} \label{sec:obs}
Comet C/2015 ER61 (PanSTARRS; hereafter ER61) is a long-period comet from the Oort cloud (\textit{P} $\sim$ 8600 yr, semi-major axis \textit{a} = 715 au) that reached perihelion on UT 2017 May 14 (\textit{q} = 1.04 au). \cite{Meech2017} reported CO$_2$-driven activity beginning near \textit{r}\subs{H} = 8.8 au, followed by H$_2$O sublimation beginning near \textit{r}\subs{H} = 5 au. \cite{Sekanina2017} reported a major outburst in ER61 beginning on UT 2017 April 3.9, followed by splitting of the nucleus, and \cite{Saki2020} reported OCS and H$_2$O production rates in ER61 on UT 2017 May 12.  We conducted pre-perihelion observations toward ER61 on UT 2017 April 11 and 15 during Cycle 4 using the ACA with the Band 7 receiver, covering frequencies between 342.3 and 364.4 GHz ($\lambda =$ 0.82 – 0.87 mm) in eight non-contiguous spectral windows ranging from 500 MHz wide (for the HCN and CO settings) to 1000 MHz wide (for the CS, HNC, CH\subs{3}OH, and H\subs{2}CO settings). The observing log is shown in Table~\ref{tab:obslog}. The comet position was tracked using JPL HORIZONS ephemerides (JPL \#43). We used two correlator settings (one on each day) to simultaneously sample spectral lines from multiple molecules and the continuum. Weather conditions were fair (precipitable water vapor at zenith; zenith PWV = 1.14 – 2.29 mm). Quasar observations were used for bandpass and phase calibration. Neptune (on April 11) and Titan (on April 15) were used to calibrate ER61’s flux scale. The spatial scale (the range in semi-major and semi-minor axes of the synthesized beam) was 2$\farcs$31 – 6$\farcs$07 and the channel spacing was 122 kHz (for HCN and CO) and 244 kHz (for CS, HNC, CH\subs{3}OH, and H\subs{2}CO), leading to a spectral resolution of 0.10 – 0.20 km\,s\sups{-1}. The data were flagged, calibrated, and imaged using standard routines in CASA version 5.6.1 \citep{McMullin2007}. We deconvolved the point-spread function with the Högbom algorithm, using natural visibility weighting and a flux threshold of twice the rms noise in each image. For HNC, a 5$\arcsec$ Gaussian \textit{uv}-taper was applied to the image to improve signal-to-noise. Self-calibration was applied to the continuum images to improve the dynamic range, resulting in an improvement from 3.4 to 11.7 on April 11 and from 4.7 to 10.2 on April 15. Self-calibration was not applied to the spectral lines. The deconvolved images were then convolved with a Gaussian fit to the PSF and corrected for the response of the ALMA primary beam. We transformed the images from astrometric coordinates to projected cometocentric distances, with the location of peak continuum flux chosen as the origin. The position of peak continuum flux (measured before the application of self-calibration) was in good agreement with the predicted ephemeris position. 

\begin{figure}[h]
\gridline{\fig{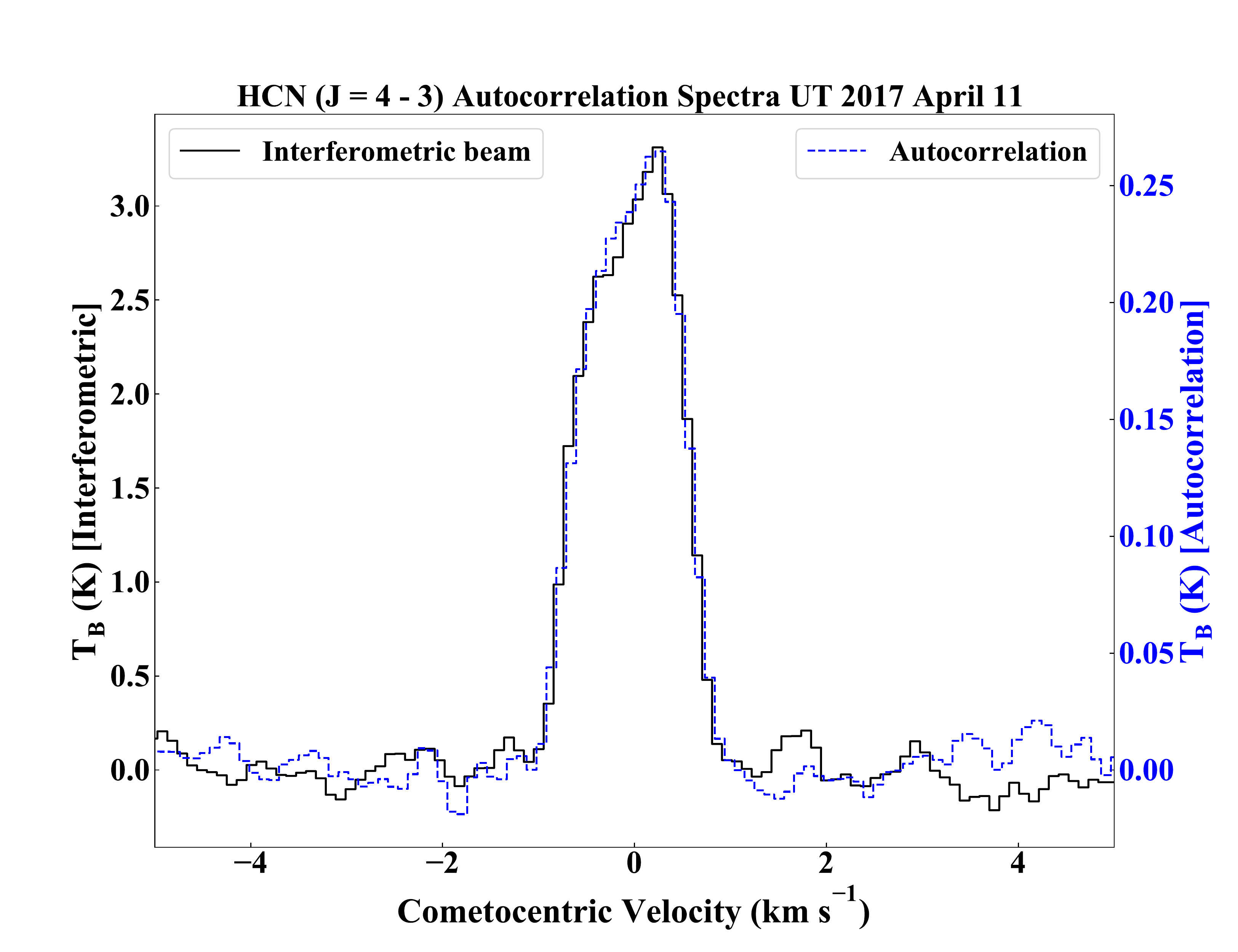}{0.45\textwidth}{(A)}
         }
\gridline{ \fig{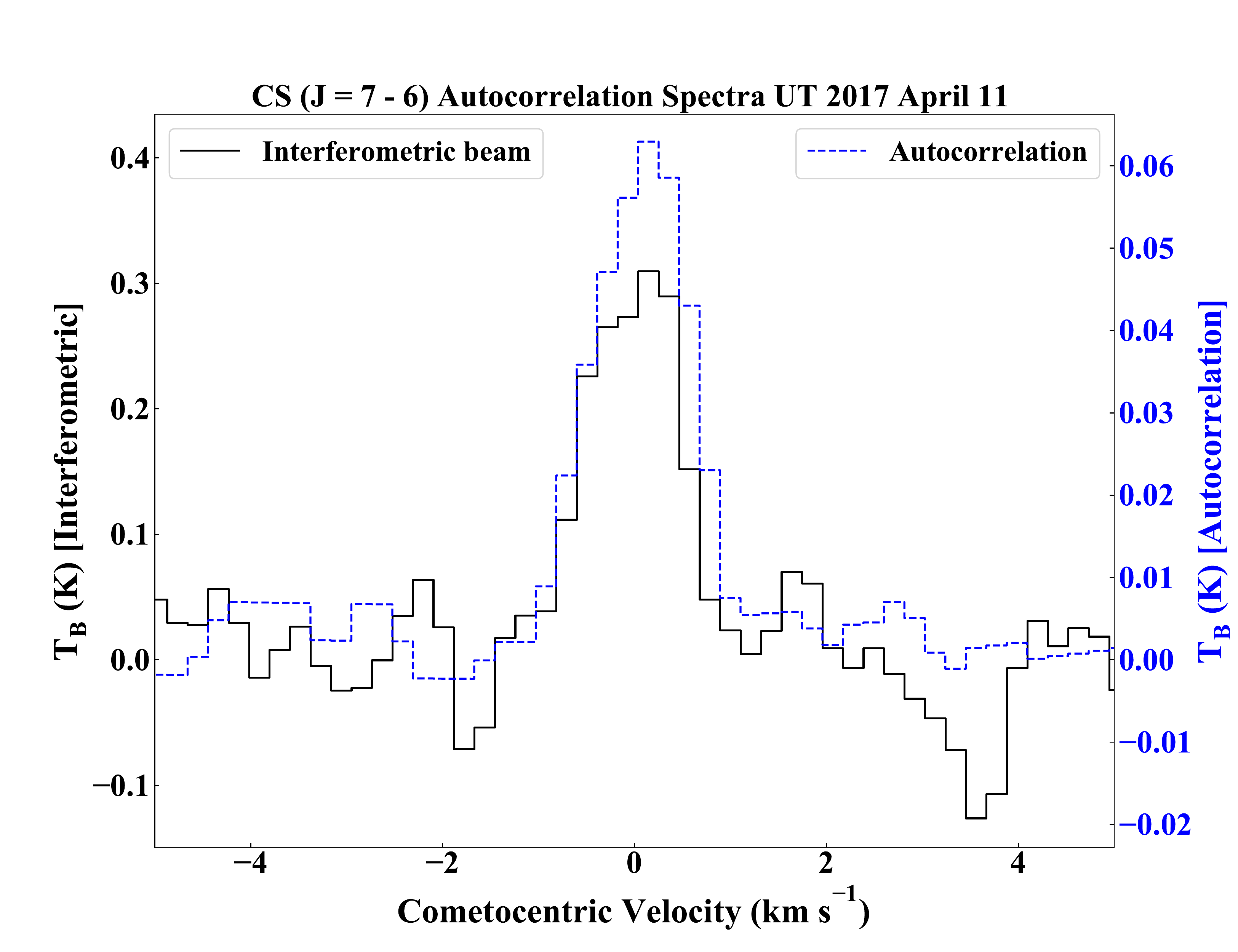}{0.45\textwidth}{(B)}
          \fig{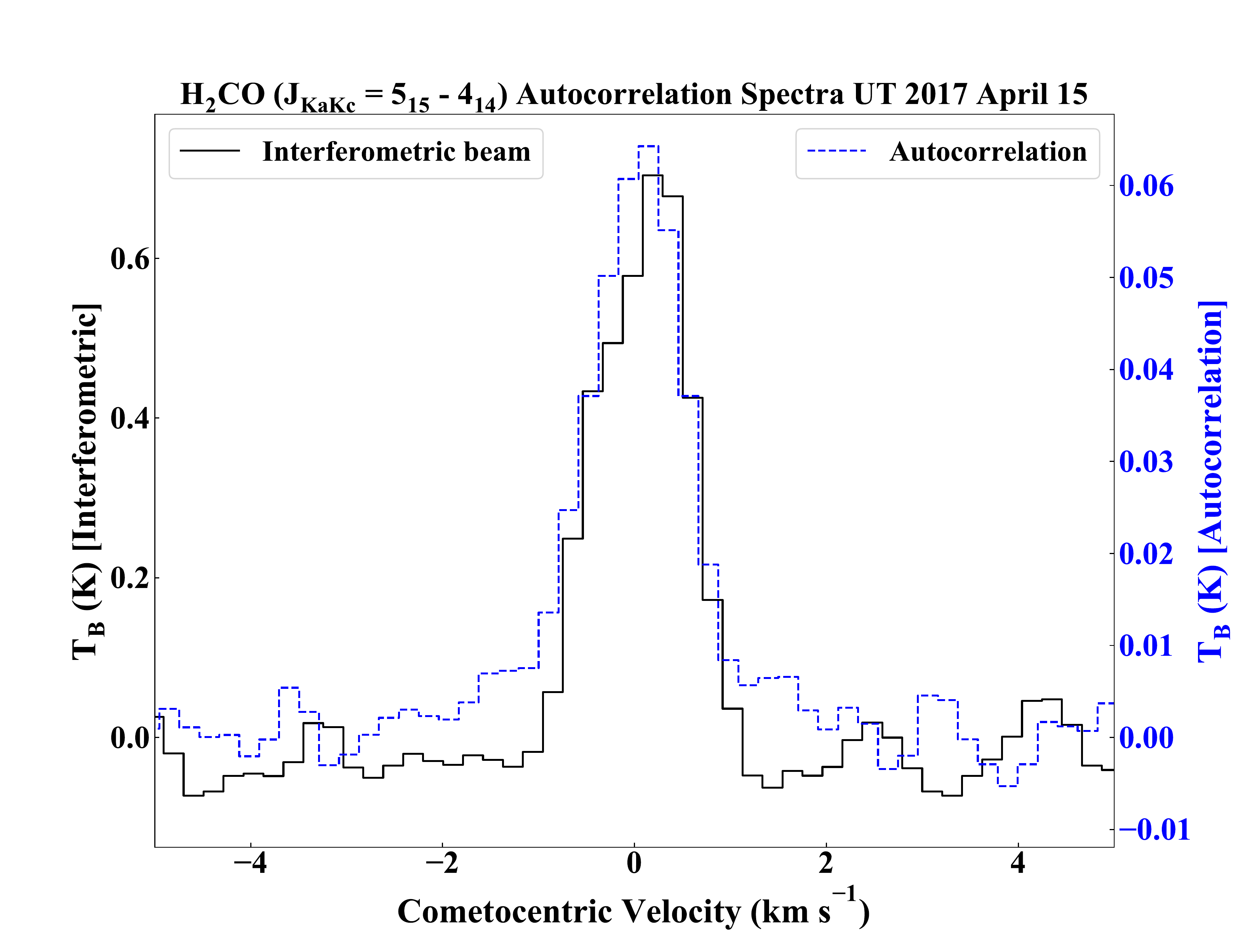}{0.45\textwidth}{(C)}
          }
\gridline{\fig{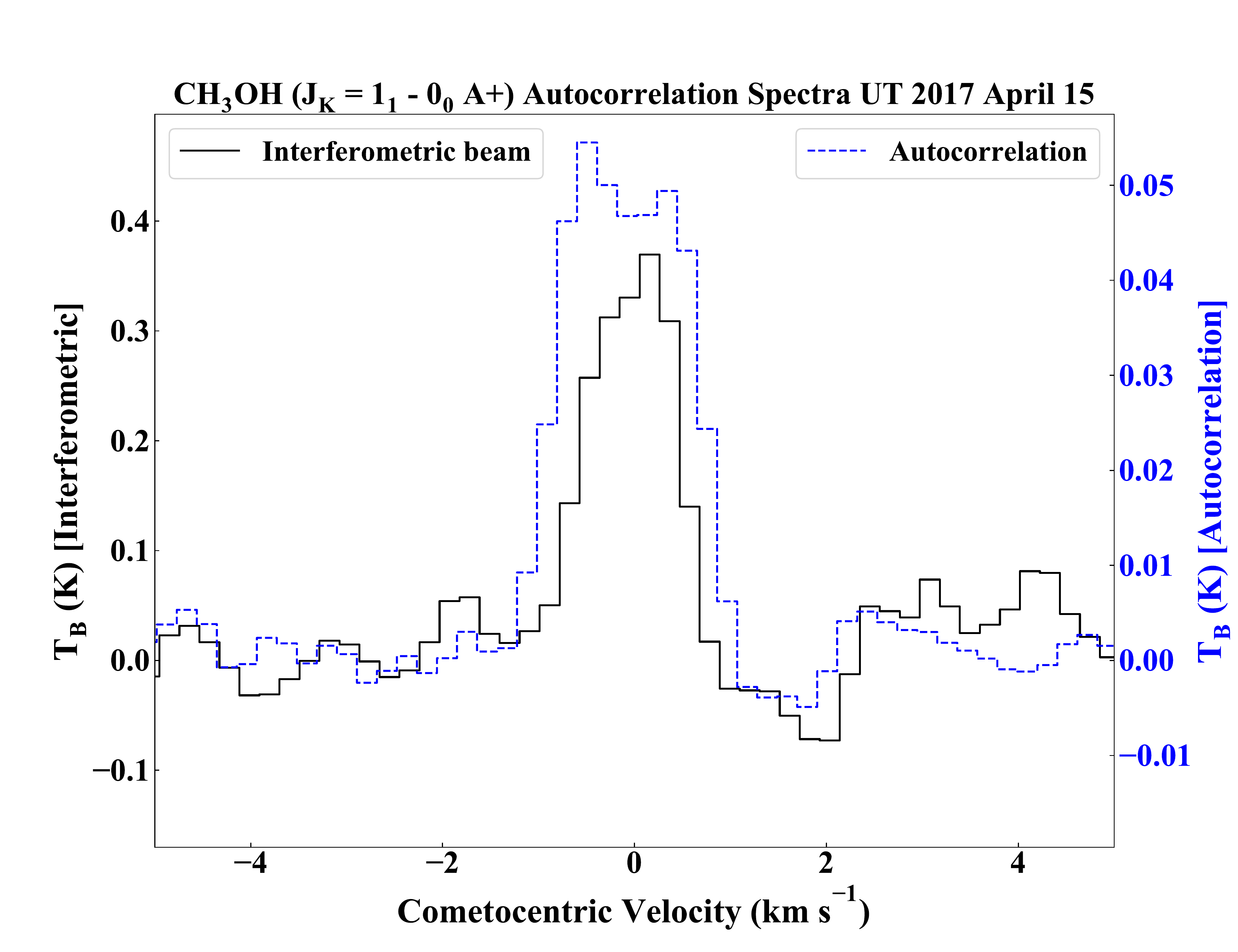}{0.45\textwidth}{(D)}
          \fig{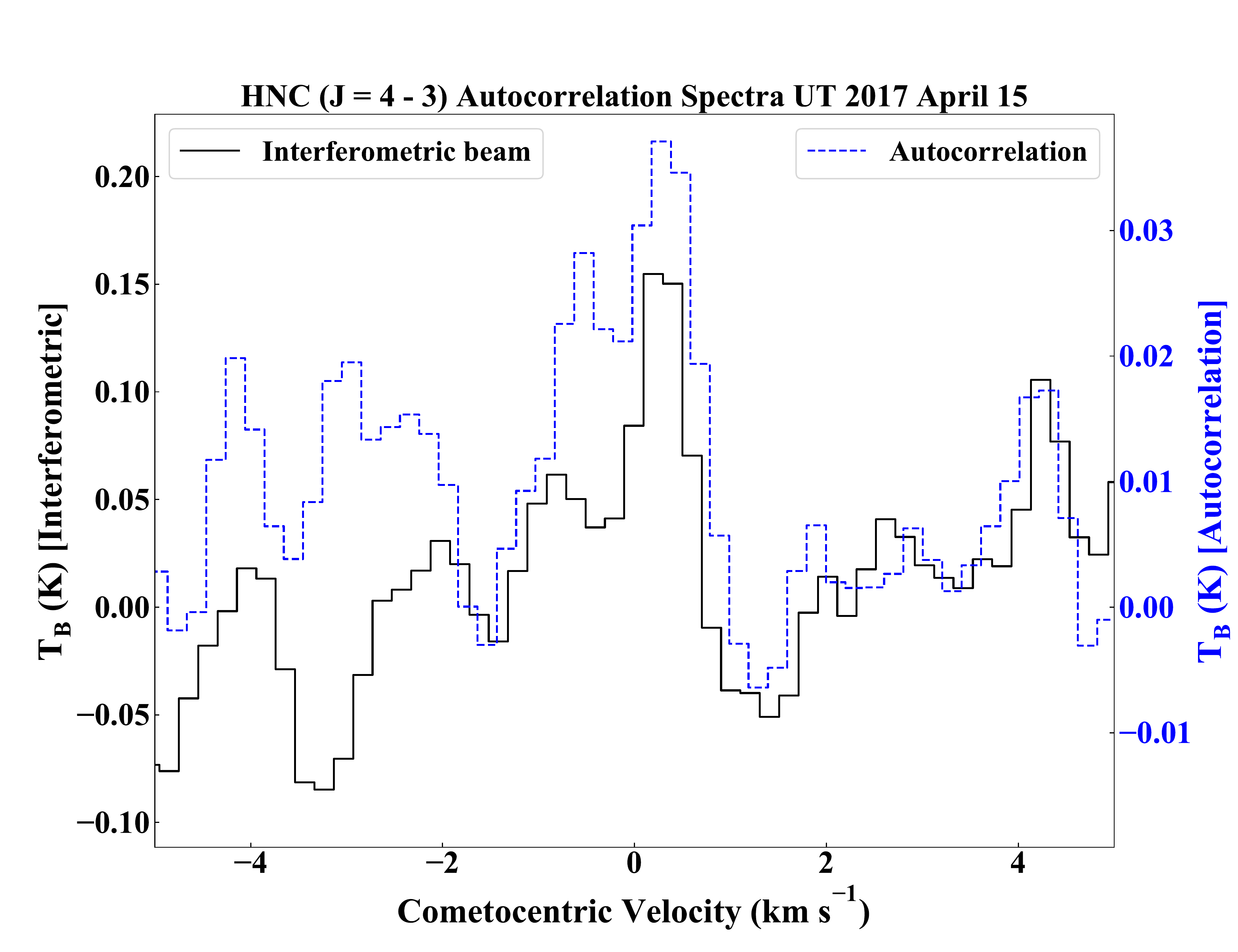}{0.45\textwidth}{(E)}
          }
\caption{\textbf{(A)--(E).} Autocorrelation spectra for HCN, CS, H$_2$CO, CH$_3$OH, and HNC in ER61 (blue dashed). Also shown are the corresponding interferometric spectra (black) for each transition. The spectral resolution is 122 kHz for HCN and 244 kHz for CS, HNC, CH$_3$OH, and H$_2$CO ($\sim$0.10--0.20 km s$^{-1}$). Note the differing scales for the \textit{y}-axis for each beam.
\label{fig:autocorr}}
\end{figure}

\subsection{Autocorrelation Spectra}\label{subsec:autocorr}
To recover extended flux that may have been resolved out by the ACA, we extracted autocorrelation spectra at the expected frequency for each transition detected in the imaging as well as for the CO (\textit{J} = 3--2) transition (not detected in imaging). These autocorrelation spectra are collected for each antenna simultaneously with the cross-correlation spectra, but are flagged (discarded) in order to process the interferometric data. We recovered, extracted, and calibrated the autocorrelation spectra on each date as in \cite{Cordiner2019,Cordiner2020}. A 20\% uncertainty in flux was assumed to account for uncertainties in the extraction process (e.g., opacity, \textit{T}\subs{sys}, background subtraction, beam efficiencies). The autocorrelation spectra were corrected for a main beam efficiency, $\eta$\subs{MB} = 0.75, as in \cite{Cordiner2020}. Figure~\ref{fig:autocorr} shows autocorrelation spectra for each detected species along with the corresponding interferometric spectra extracted from the nucleus-centered beam. 

\section{Results}\label{sec:results}
We detected molecular emission from HCN, CS, CH\subs{3}OH, H\subs{2}CO, and HNC as well as continuum emission from dust/icy grains in the coma of ER61, and derived a sensitive upper limit for CO. We modeled molecular line emission using a three-dimensional radiative transfer method based on the Line Modeling Engine \citep[LIME;][]{Brinch2010} adapted for cometary atmospheres, including a full non-LTE treatment of coma gases, collisions with H\subs{2}O and electrons, and pumping by solar radiation \citep[Appendix~\ref{sec:pumping}; see][for further details]{Cordiner2019}. Our models were run through the CASA ``simobserve'' tool to mimic the effects of the ACA using the exact antenna configuration and hour angle range of our observations. Photodissociation rates for all molecules were adopted from \cite{Huebner2015} with the exception of CS \citep{Boissier2007}. Detected spectral lines, including upper-state energies (\textit{E}\subs{u}), calculated production rates (\textit{Q}), parent scale lengths (\textit{L}\subs{p}), and integrated fluxes are listed in Table~\ref{tab:comp}.

\begin{deluxetable*}{ccccccccc}[h]
\tablenum{2}
\tablecaption{Molecular Composition of C/2015 ER61 (PanSTARRS)\label{tab:comp}}
\tablewidth{0pt}
\tablehead{
\colhead{Species} & \colhead{Transition} & \colhead{\textit{E}\subs{u}} & \colhead{$\int T_{B} \,dv$\sups{a}} & \colhead{$\int T_B \,dv$\sups{b}} & \colhead{\textit{L}\subs{p}\sups{c}}  & \colhead{\textit{Q}\sups{d}} &  \colhead{\textit{Q}\subs{x}/\textit{Q}\subs{H2O}\sups{e}} & \colhead{Range in Comets\sups{f}} \\
\colhead{} & \colhead{} & \colhead{(K)} & \colhead{(K km s$^{-1}$)} & \colhead{(K km s$^{-1}$)} & \colhead{(km)} &  \colhead{(10\sups{25} mol s$^{-1}$)} &  \colhead{(\%)} & \colhead{(\%)}
}
\startdata
\multicolumn{9}{c}{UT 2017 April 11, \textit{r}\subs{H} = 1.14 au, $\Delta$ = 1.19 au} \\
\hline
HCN & 4--3 & 25.5 & 3.69 $\pm$ 0.06 & 0.26 $\pm$ 0.05 & $50^{+50}_{-30}$ &  8.6 $\pm$ 0.3 & 0.072 & 0.08--0.25\\
CS    & 7--6 & 65.8 & 0.25 $\pm$ 0.05 & 0.082 $\pm$ 0.016 & $2000^{+3100}_{-1200}$ &   7.4 $\pm$ 0.9 &  0.061 & 0.02--0.20\\
CO   & 3--2 & 33.2 & $<$ 0.147 (3$\sigma$)   & $<$ 0.009 (3$\sigma$)   & $\cdot \cdot \cdot$ &  $<$ 191 (3$\sigma$) & $<$ 1.6 (3$\sigma$) & 0.3--23\\
\hline
\multicolumn{9}{c}{UT 2017 April 15, \textit{r}\subs{H} = 1.12 au, $\Delta$ = 1.18 au} \\
\hline
H\subs{2}CO & $5_{15}$--$4_{14}$ & 62.4 & 0.58 $\pm$ 0.04 & 0.085 $\pm$ 0.017 & $2200^{+1100}_{-800}$ &  37 $\pm$ 3  & 0.31 & 0.13-1.4 \\
CH\subs{3}OH & $1_1$--$0_0$ $A^{+}$  & 16.8 & 0.40 $\pm$ 0.05 &  0.084 $\pm$ 0.017 & $300^{+1400}_{-260}$ &  230 $\pm$ 88 & 1.9 & 0.6--6.2 \\
HNC & 4--3 & 43.5 & 0.36 $\pm$ 0.07 & 0.046 $\pm$ 0.009 & $3300^{+19,700}_{-2800}$ &  0.82 $\pm$ 0.17 & 0.007 & 0.002--0.035
\enddata
\tablecomments{\sups{a} Spectrally integrated flux extracted from the interferometric nucleus-centered beam at the position of peak continuum flux. \sups{b} Spectrally integrated flux extracted from autocorrelation spectra. \sups{c} Parent scale length derived from interferometric visibility modeling. \sups{d} Molecular production rate calculated using the best-fit parent scale length. \sups{e} Mixing ratio with respect to H$_2$O with \textit{Q}(H$_2$O) = 1.2 $\times 10^{29}$ s$^{-1}$ as measured by \cite{Saki2021} on UT 2017 April 15. \sups{f} Range of mixing ratios in comets \citep{DelloRusso2016a,Bockelee-Morvan2017}.
}
\end{deluxetable*}

\subsection{Interferometric Molecular Maps, Extracted Spectra}\label{subsec:maps}
\begin{figure*}
\gridline{\fig{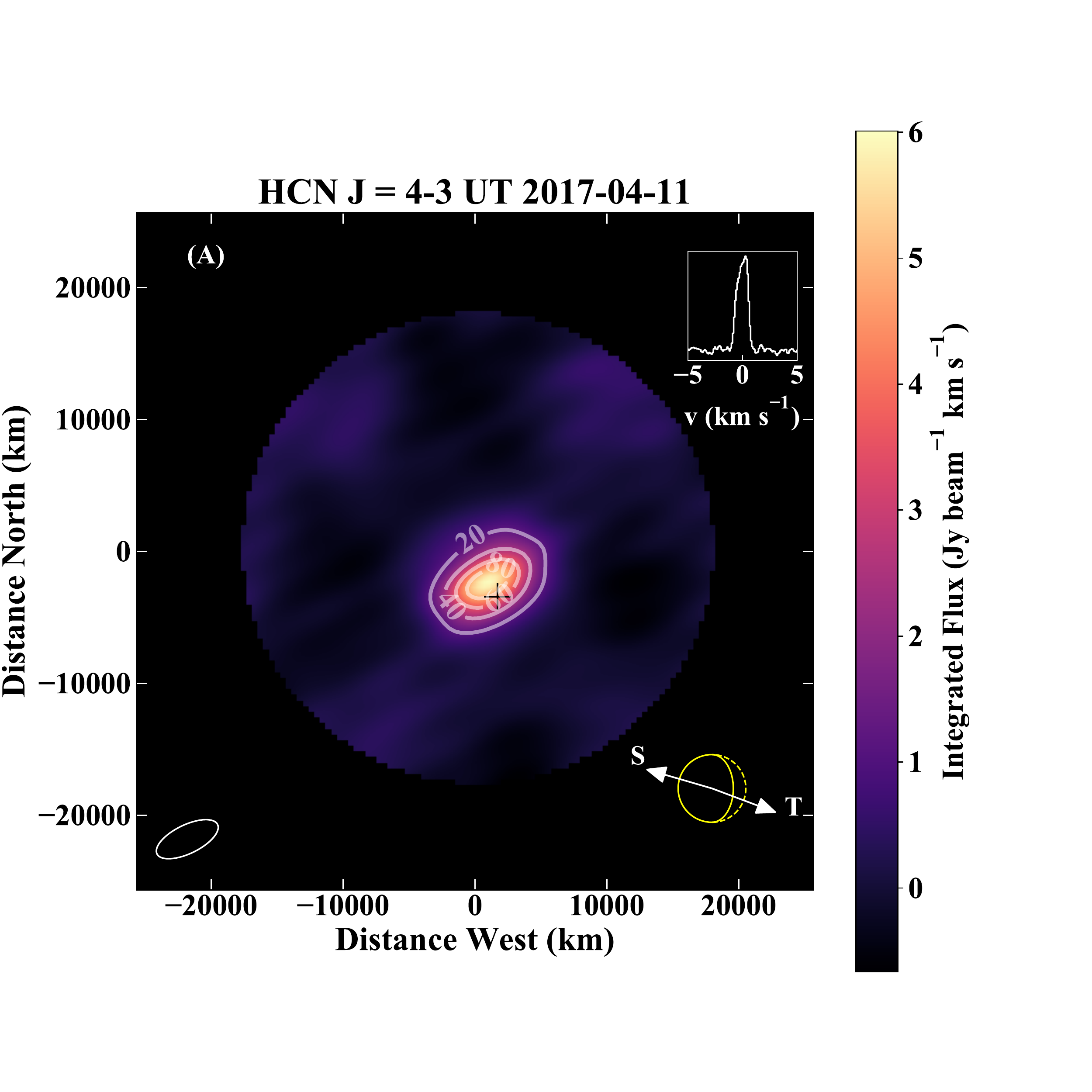}{0.5\textwidth}{(A)}
          \fig{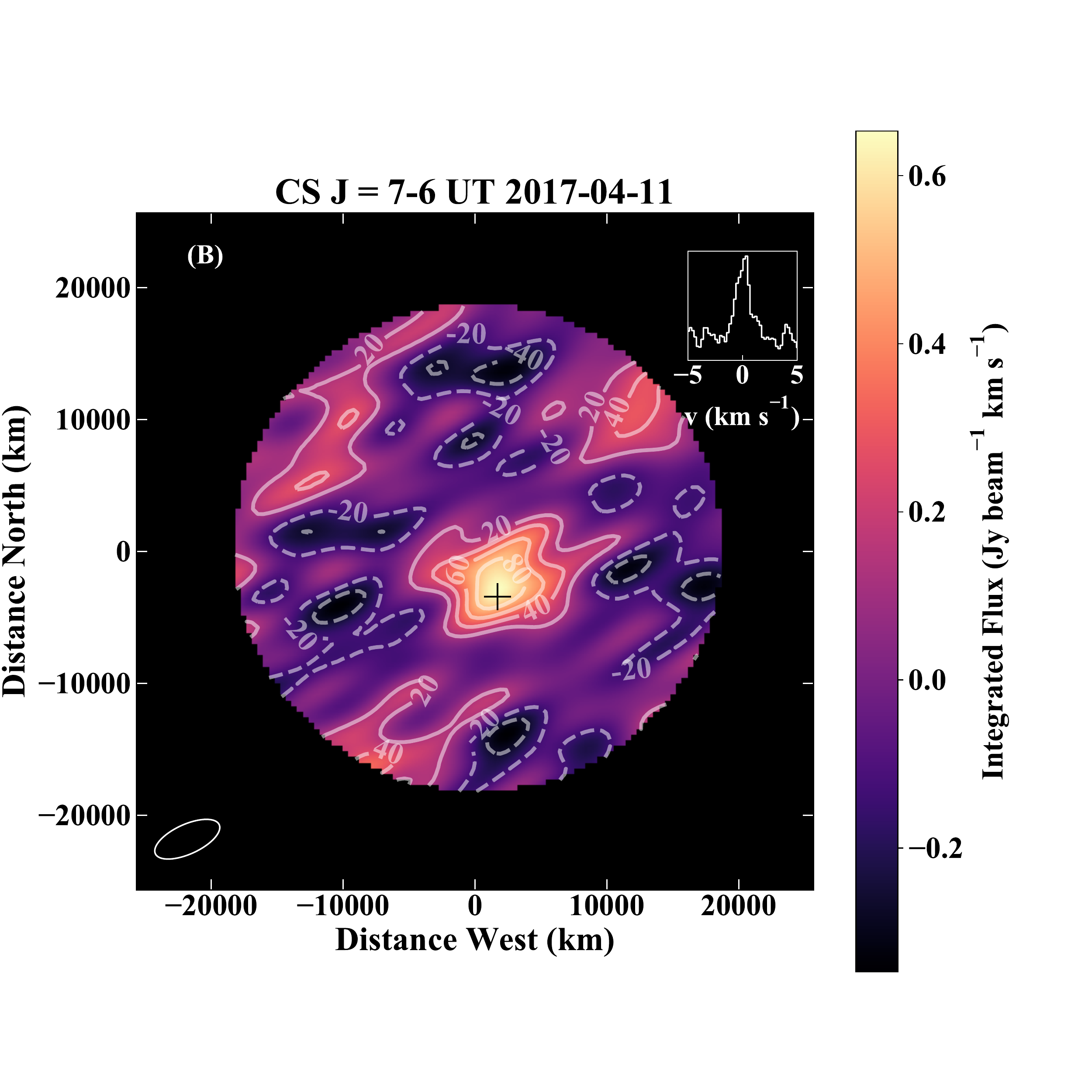}{0.5\textwidth}{B}
	}
\gridline{\fig{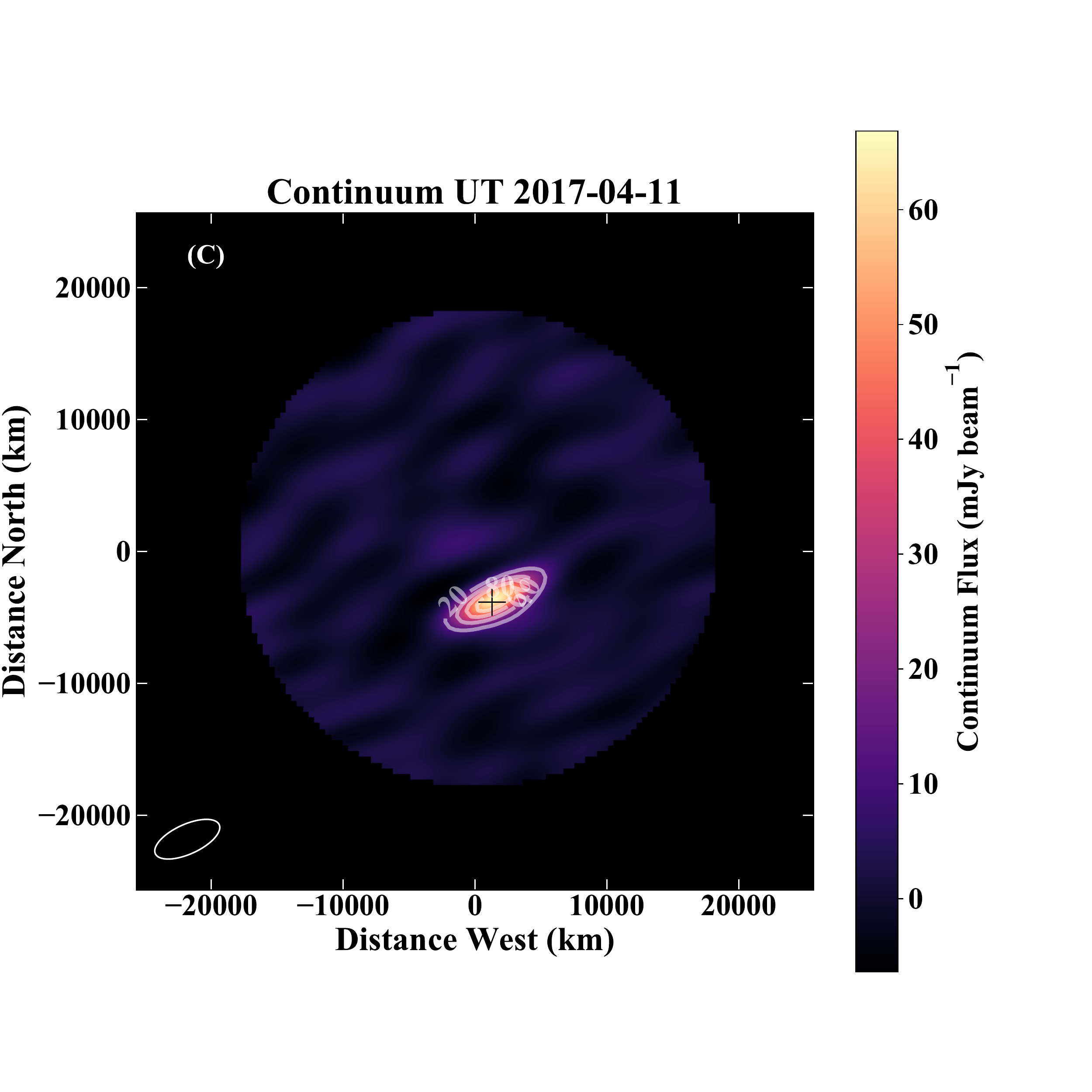}{0.5\textwidth}{(C)}
	}
\caption{\textbf{(A)--(C).} Spectrally integrated flux maps for HCN, CS, and continuum in ER61 on UT 2017 April 11.
Contour intervals in each map are 20\% of the peak. The rms noise ($\sigma$, mJy beam$^{-1}$ km s$^{-1}$) and contour spacing, $\delta$, for each map are: (A) HCN: $\sigma$ = 230, $\delta$ = 5.2$\sigma$, (B) CS: $\sigma$ = 120, $\delta$ = 1.1$\sigma$, and (C) Continuum: $\sigma$ = 4.4 mJy beam$^{-1}$, $\delta$ = 3.1$\sigma$. Sizes and orientations of the synthesized beam (6$\farcs$07$\times$2$\farcs$56) are indicated in the lower left corner of each panel. The comet's illumination phase ($\phi \sim$ 51$\degr$), as well as the direction of the Sun and dust trail, are indicated in the lower right. The black cross indicates the position of the peak continuum flux. A spectral line profile taken at the peak of emission is shown in the upper right.
\label{fig:maps1}}
\end{figure*}

\begin{figure*}
\gridline{
          \fig{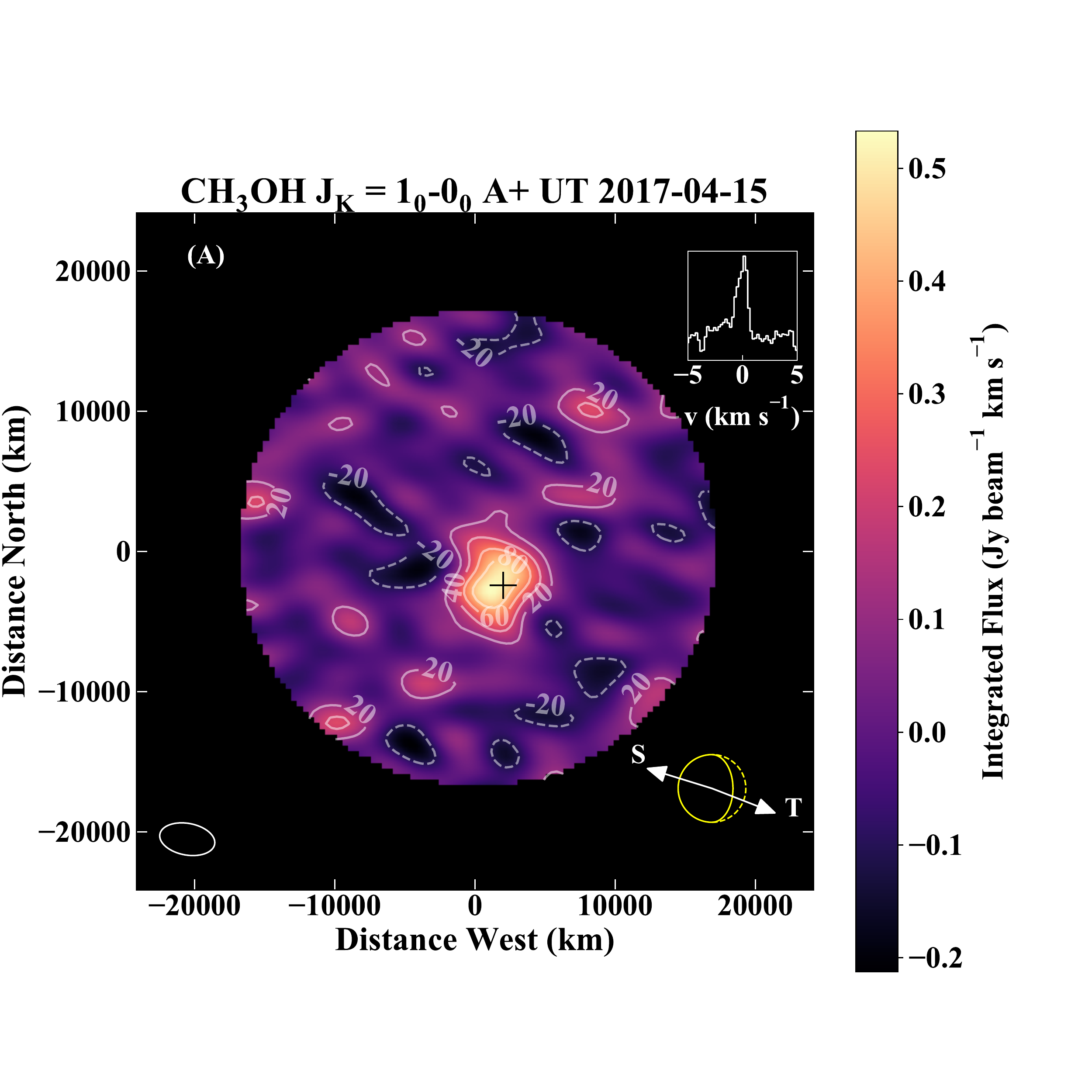}{0.5\textwidth}{(A)}
          \fig{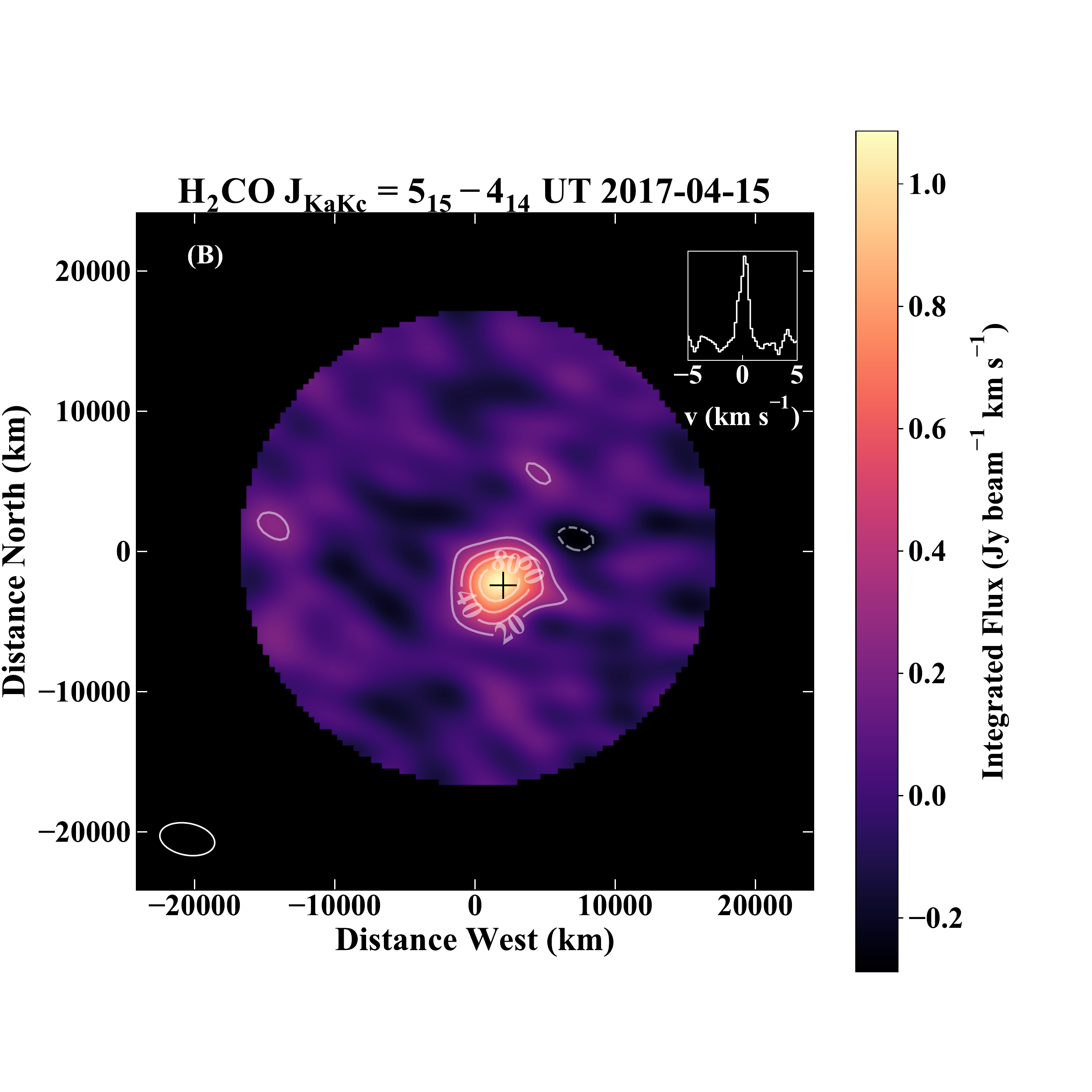}{0.5\textwidth}{(B)}
          }
\gridline{
          \fig{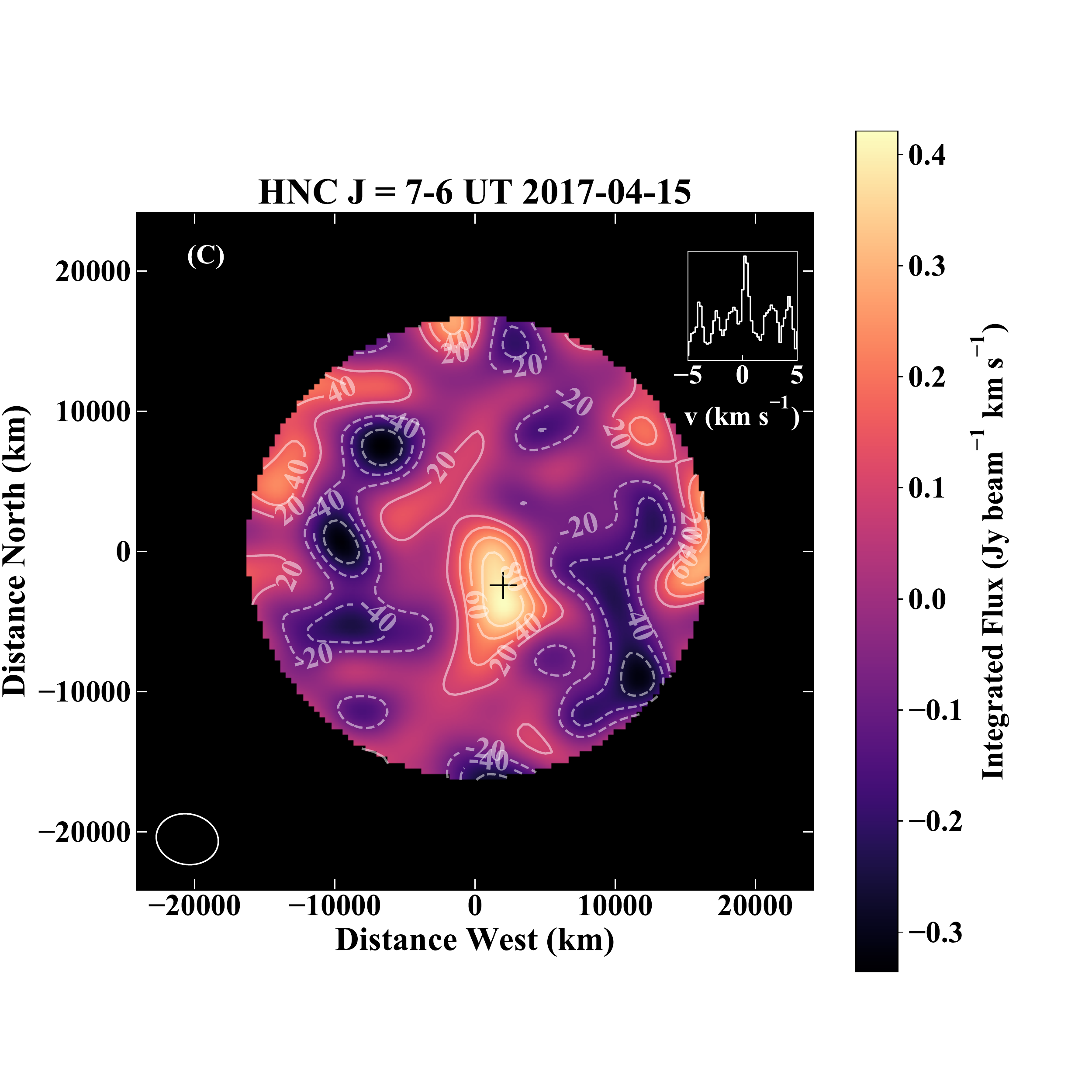}{0.5\textwidth}{(C)}
          \fig{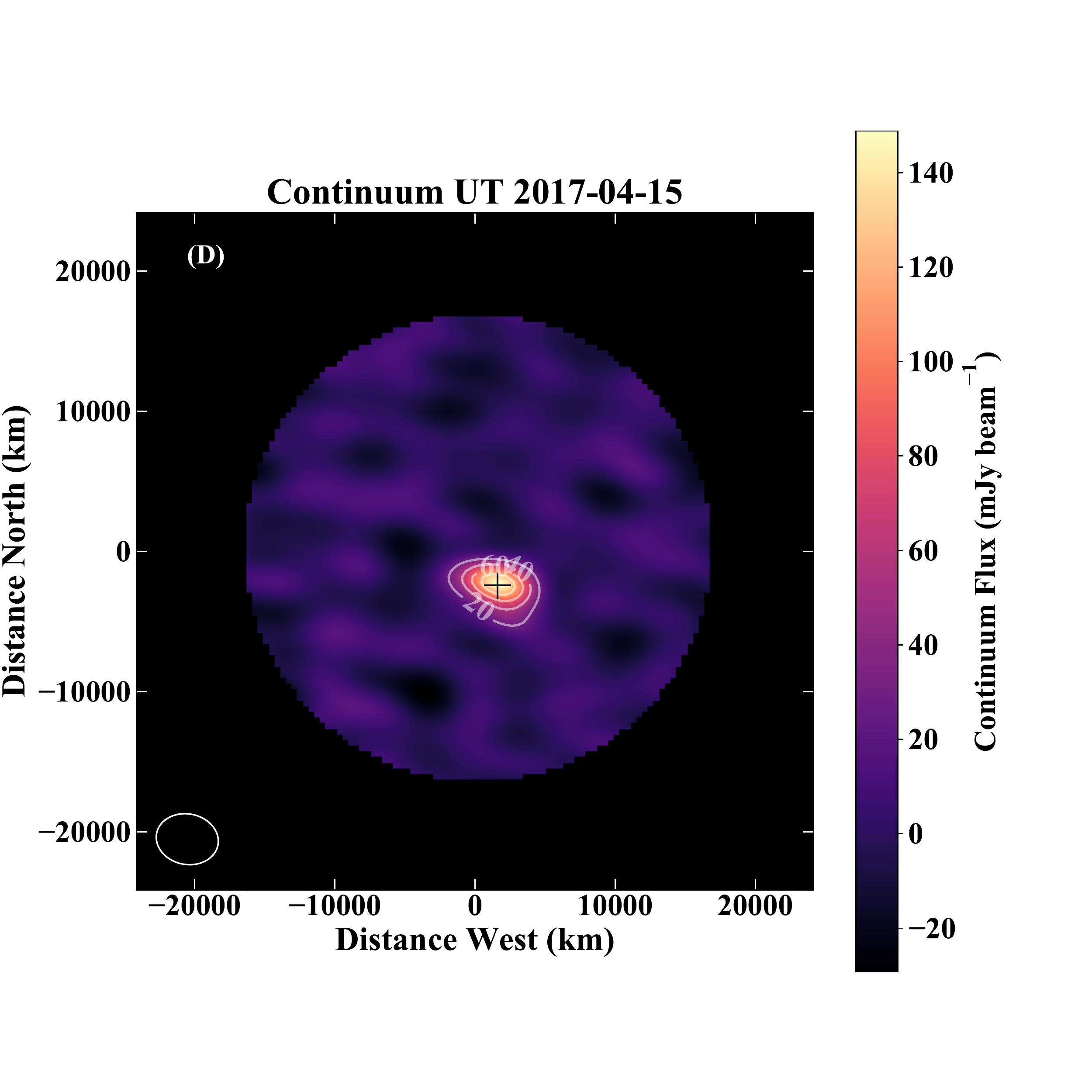}{0.5\textwidth}{(D)}
           }
\caption{\textbf{(A)--(D).} Spectrally integrated flux maps for CH$_3$OH, H$_2$CO, HNC, and continuum on UT 2017 April 15, with traces and labels as in Figure~\ref{fig:maps1}.
Contour intervals in each map are 20\% of the peak. The rms noise ($\sigma$, mJy beam$^{-1}$ km s$^{-1}$) and contour spacing, $\delta$, for each map are: (A) CH$_3$OH: $\sigma$ = 69, $\delta$ = 1.9$\sigma$, (B) H$_2$CO: $\sigma$ = 49, $\delta$ = 4.3$\sigma$, (C) HNC: $\sigma$ = 138, $\delta$ = 0.61$\sigma$, and (D) Continuum: $\sigma$ = 3.3 mJy beam$^{-1}$, $\delta$ = 9.0$\sigma$. A spectral line profile taken at the peak of emission is shown in the upper right.
\label{fig:maps2}}
\end{figure*}

Figures~\ref{fig:maps1} and~\ref{fig:maps2} show observed spectra and spectrally integrated flux maps for each species detected with the ACA in ER61. Differences are evident between the emissions of individual molecules as well as dust. ER61 displays a compact dust continuum that falls off within a few thousand km of the nucleus, whereas molecular emission extends up to greater nucleocentric distances. The distribution of HCN is compact and consistent with that expected for a parent species, whereas those of CS and HNC are more extended and consistent with a product or distributed source distribution. On April 11, the continuum peak is offset from the HCN peak by $\sim$1200 km to the southwest (i.e., anti-sunward), whereas on April 15 the continuum and molecular emission peaks are better aligned, with the continuum peaking $\sim$850 km to the northeast (sunward) of the H$_2$CO emission. The continuum emission is unresolved and further discussed in Section~\ref{sec:cont}. We performed a detailed analysis of the parentage of each species (Section~\ref{sec:prod}). Additionally, the extracted spectra are all indicative of a red-shifted peak, suggesting asymmetries in outgassing along the Sun--comet line (Section~\ref{subsec:vexp}).

\begin{figure*}
\gridline{\fig{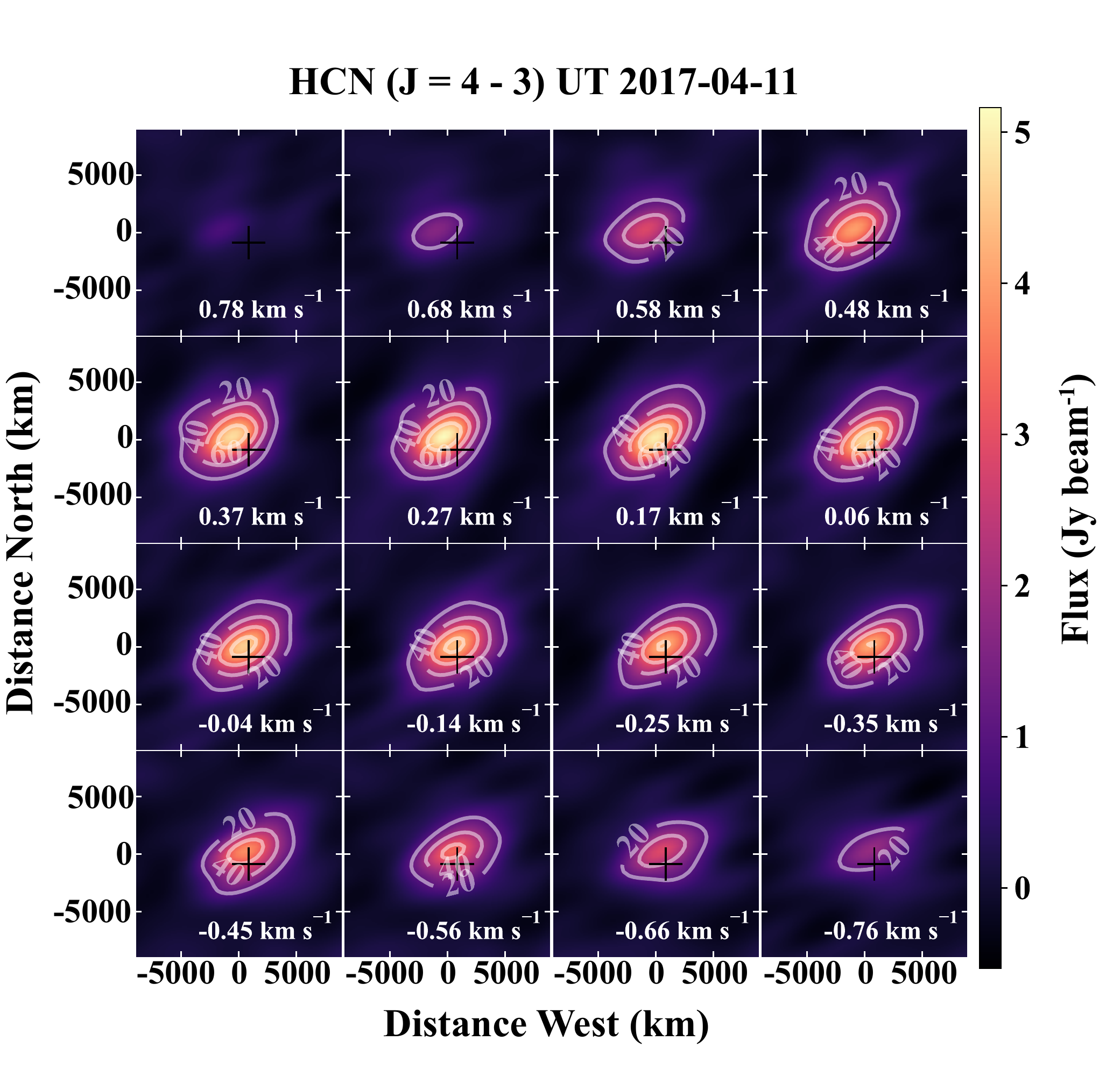}{0.5\textwidth}{(A)}
	}
\gridline{\fig{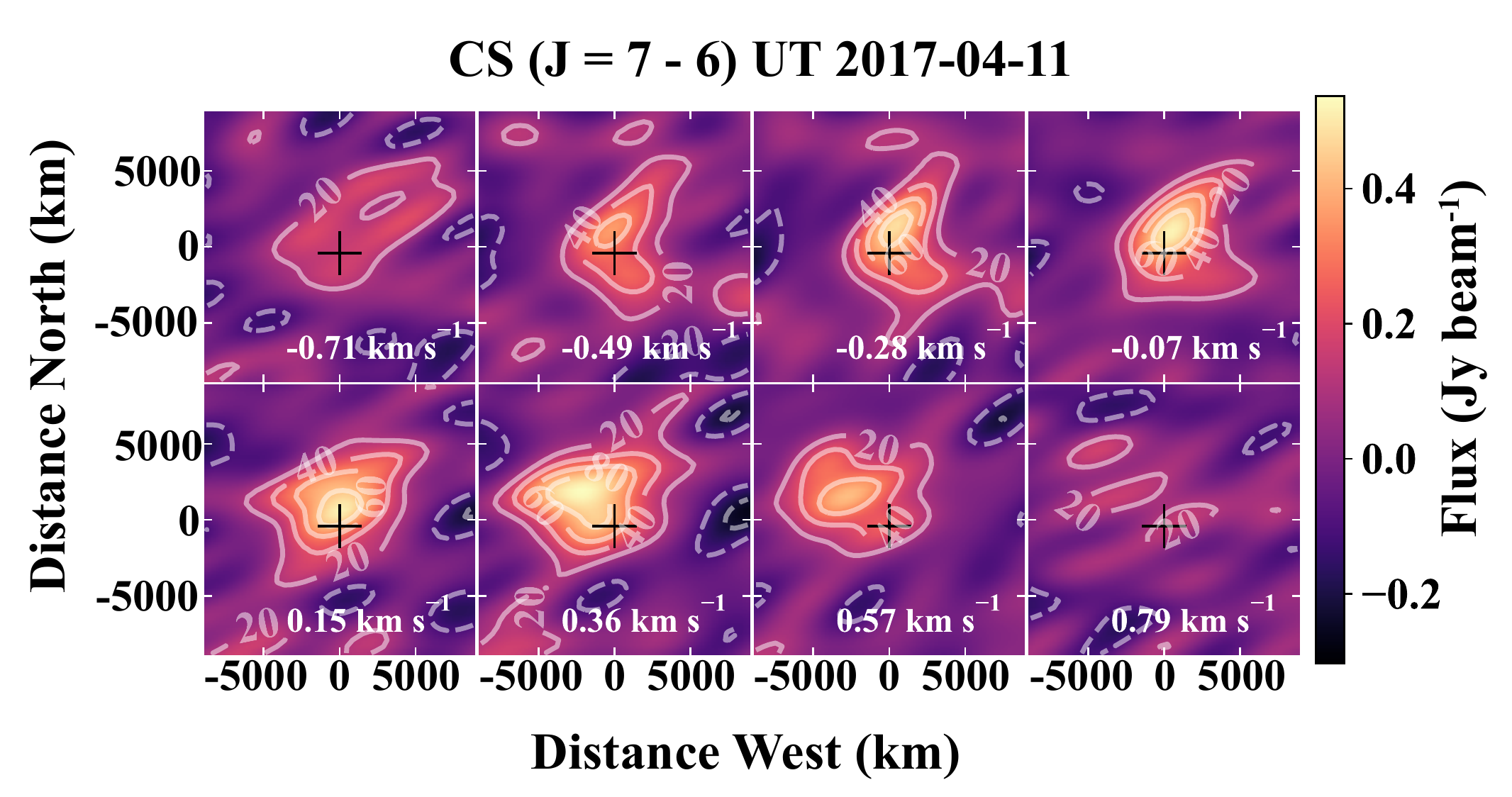}{0.5\textwidth}{(B)}
          \fig{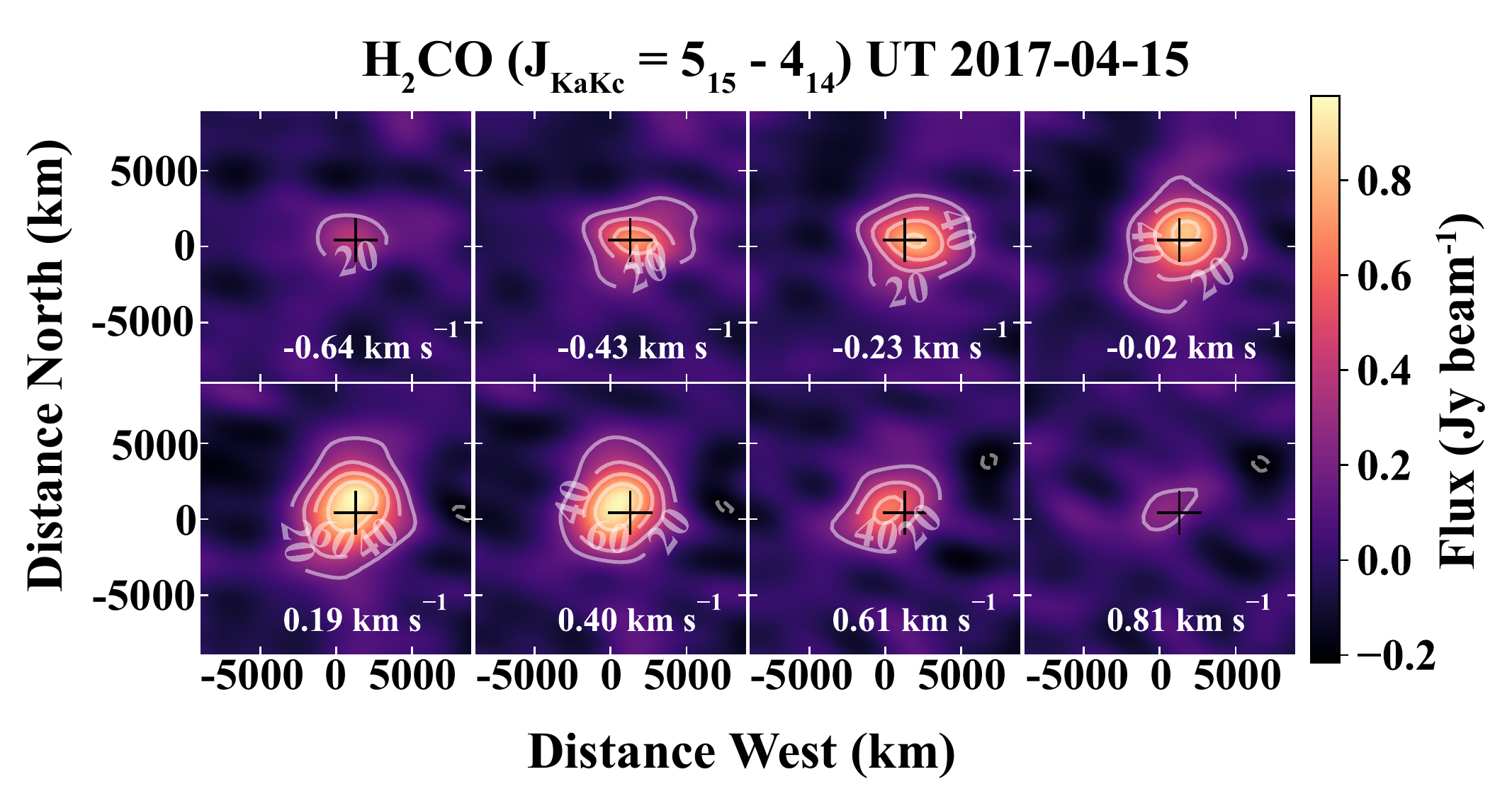}{0.5\textwidth}{(C)}
          }
\gridline{\fig{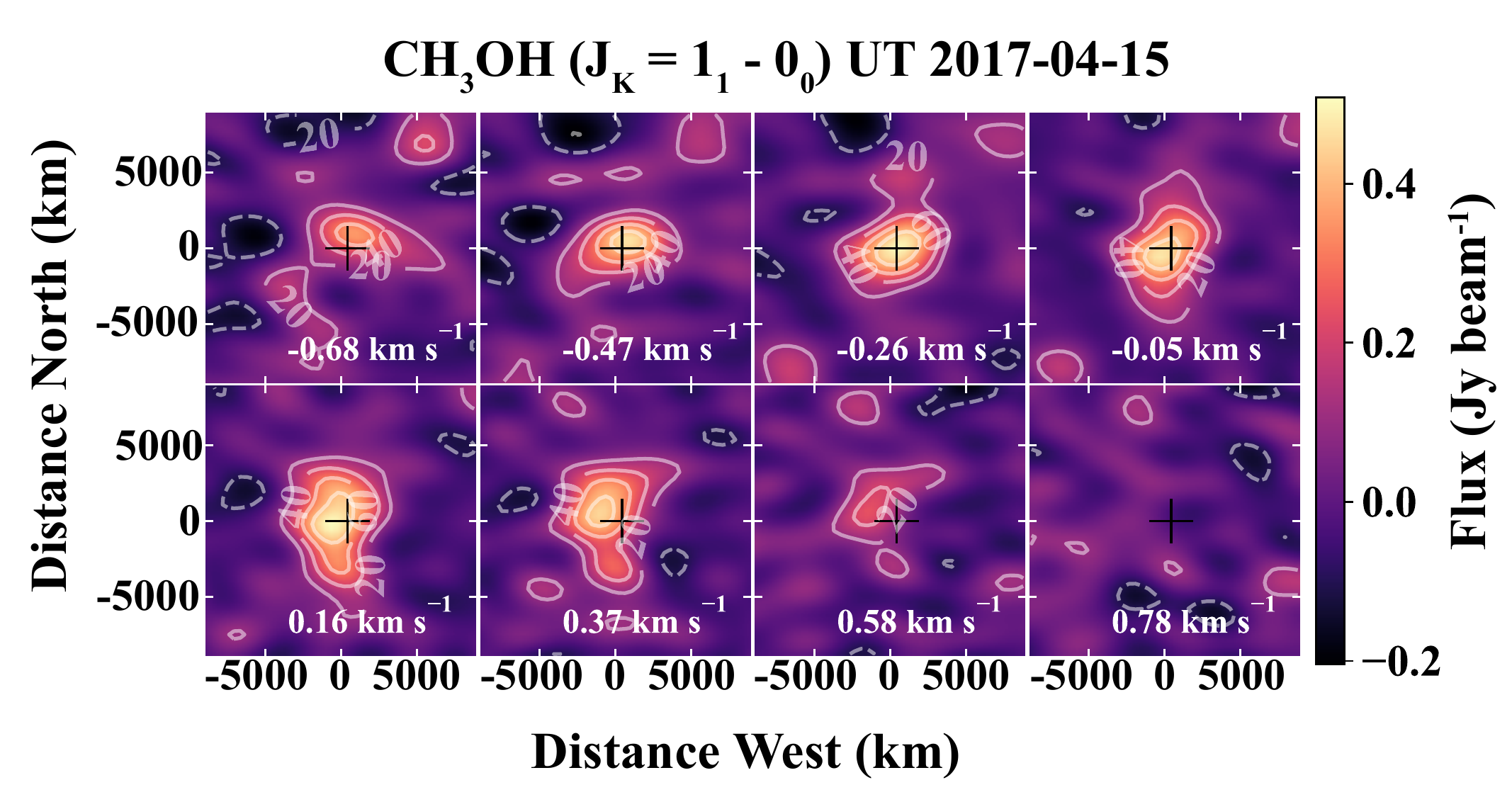}{0.5\textwidth}{(D)}
          \fig{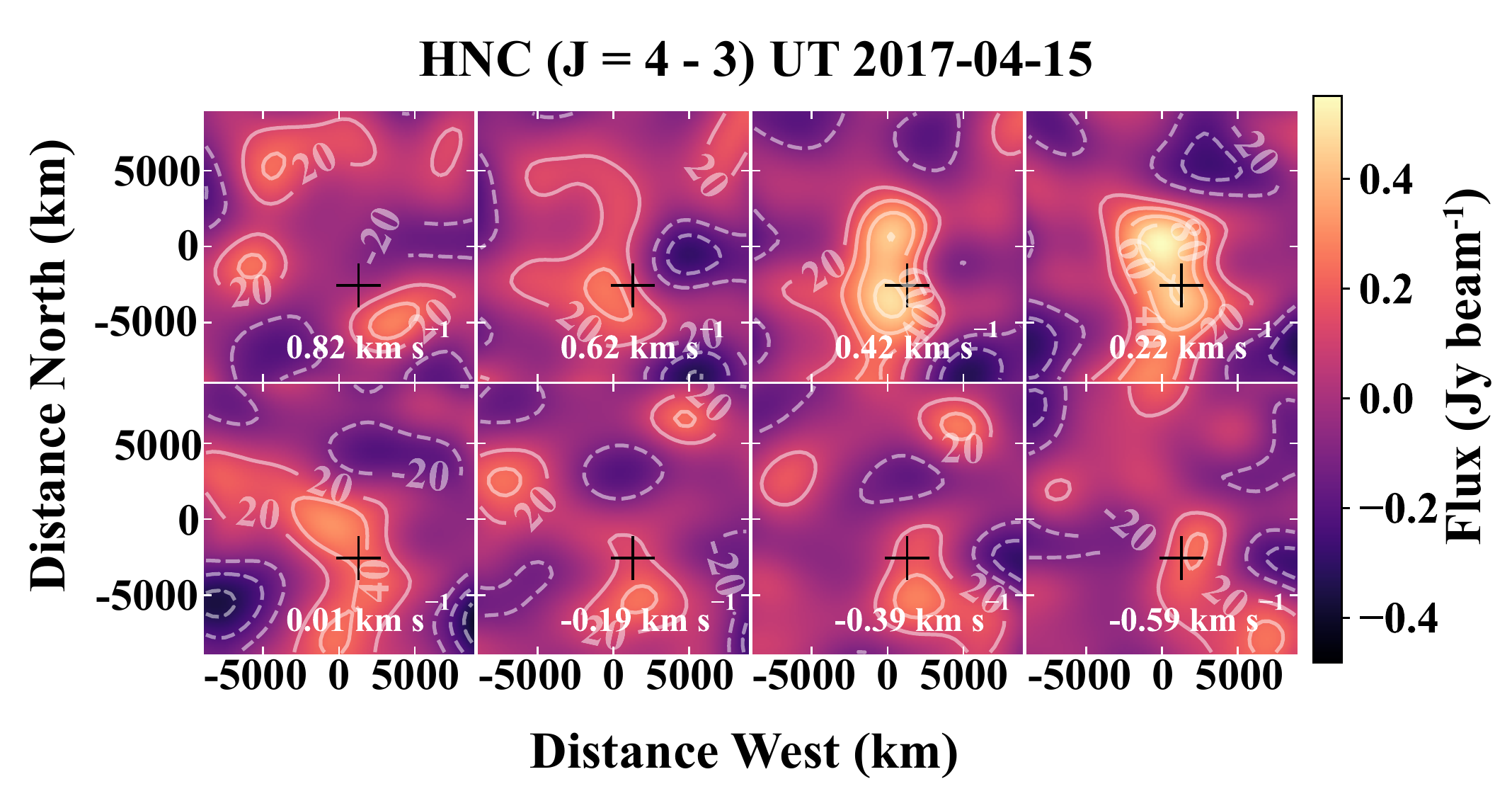}{0.5\textwidth}{(E)}
          }
\caption{\textbf{(A)--(E).} Velocity channel maps for (A) HCN, (B) CS, (C) H\subs{2}CO, (D) CH\subs{3}OH, and (D) HNC in ER61. Each contour level is 20\% of the peak flux across all channels. The cometocentric velocity of each channel is indicated in the bottom for each map, and the peak continuum flux position is indicated by a black cross.
\label{fig:chanmaps}}
\end{figure*}

The complexity of ER61's coma as captured by the ACA is further revealed in velocity channel maps for each molecule (Figure~\ref{fig:chanmaps}). The spatial extents of emissions for each molecule are not constant across velocity space. Whereas the overall shape of HCN and H\subs{2}CO emissions are relatively consistent across a range of cometocentric velocities, CS and HNC (and to a lesser extent, CH\subs{3}OH) exhibit asymmetric structures varying in extent and shape across a relatively narrow range of velocity space.

%The azimuthally averaged radial profiles of each molecule (Figure~\ref{fig:radprof}) can provide further insight into their heritage in the coma. These profiles trace the radial intensity of molecular emission as a function of nucleocentric distance for each of the (optically thin) transitions in Table~\ref{tab:comp}. The continuum radial profiles trace that for a point-source on each date, consistent with relatively compact emission from dust (and possibly, the nucleus itself). Similarly, the radial profile of HCN falls off with the $\sim$1/$\rho$ dependence expected for volatiles that are subliming directly from the nucleus with a constant outflow velocity (where $\rho$ is the nucleocentric distance). In contrast, the profiles of CS and HNC decay considerably slower and are consistent with product species in the coma. The profiles of H$_2$CO \citep[often associated with distributed sources; e.g.,][]{Cordiner2014} and CH$_3$OH closely track one another (although each still falls off more rapidly than HNC), which may indicate distributed source production of CH$_3$OH in ER61 (Section~\ref{subsec:ch3oh}).

%\begin{figure}
%\plotone{Figure3.pdf}
%\caption{\textbf{(A)--(B).} Azimuthally averaged radial profiles for HCN, CS, H\subs{2}CO, CH\subs{3}OH, and HNC in ER61. Also shown are profiles for the dust continuum and a point-source (bandpass calibrator) on each date. All profiles are normalized to unity. \label{fig:radprof}}
%\end{figure}

\subsection{Rotational and Temperature}\label{subsec:trot}
\begin{figure}
\plotone{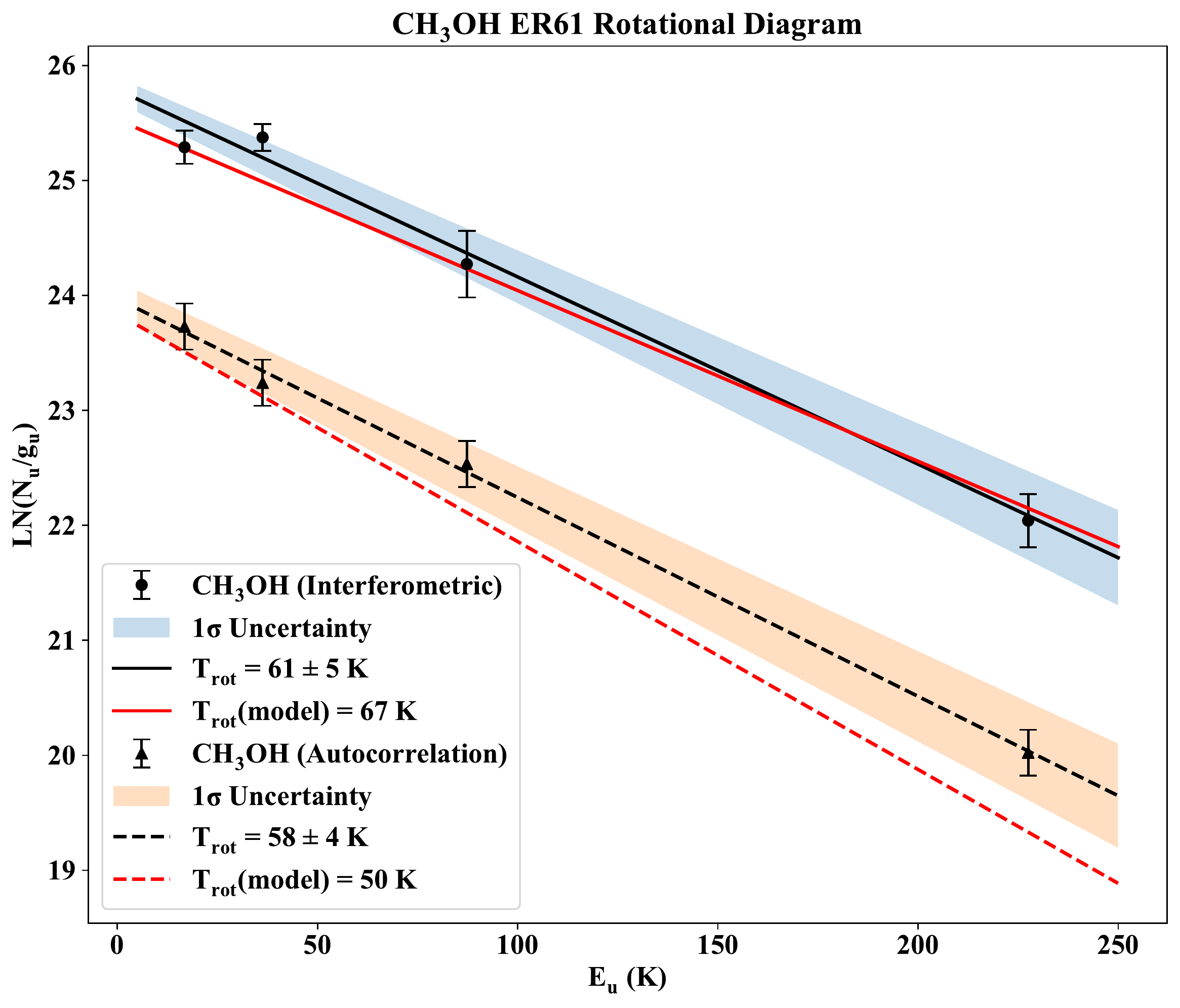}
\caption{Rotational diagram of the CH\subs{3}OH \textit{J}\subs{K} = $13_1$--$13_0$ $A^{-+}$ transition in ER61 on UT 2017 April 11 and the \textit{J}\subs{K} = $1_1$--$0_0$ $A^{+}$, \textit{J}\subs{K} = $4_0$--$3_{-1}$ $E$, and \textit{J}\subs{K} = $7_2$--$6_1$ $E$ transitions on UT 2017 April 15 for interferometric nucleus-centered (black circles) and autocorrelation (black triangles) spectra. The best-fit lines, corresponding to \textit{T}\subs{rot} = 61 $\pm$ 5 K for the interferometric spectra and 58 $\pm$ 4 K for the autocorrelation spectra are shown as well as the 1$\sigma$ uncertainty in the fits (shaded areas). Also shown are the rotational diagrams for the best-fit non-LTE models for \textit{T}\subs{kin} = 75 K (interferometric spectra, solid red line) and for \textit{T}\subs{kin} = 88 K (autocorrelation spectra, dashed red line). Note that the interferometric data are extracted from the nucleus-centered beam whereas the autocorrelation data are from the considerably larger autocorrelation beam (Table~\ref{tab:obslog}). \label{fig:rotdiagram}}
\end{figure}

We detected emission from one CH\subs{3}OH transition on April 11 and three additional CH\subs{3}OH transitions on April 15. We constrained the coma rotational temperature along the line of sight using the rotational diagram method \citep{Bockelee1994} for both the interferometric and autocorrelation spectra. Our results indicate a best-fit rotational temperature \textit{T}\subs{rot} = 61 $\pm$ 5 K for the interferometric nucleus-centered spectra and \textit{T}\subs{rot} = 58 $\pm$ 4 K for the autocorrelation spectra. Table~\ref{tab:trot} provides line transitions and integrated fluxes for each CH\subs{3}OH line used in the analysis and Figure~\ref{fig:rotdiagram} shows our CH\subs{3}OH rotational diagrams and best-fit lines.

\begin{deluxetable*}{ccccc}
\tablenum{3}
\tablecaption{CH\subs{3}OH Lines Detected in C/2015 ER61 (PanSTARRS) \label{tab:trot}}
\tablewidth{0pt}
\tablehead{
\colhead{Transition} & \colhead{Frequency} & \colhead{\textit{E}\subs{u}} & \colhead{$\int T_{B} \,dv$\sups{a}} & \colhead{$\int T_B\,dv$\sups{b}} \\
\colhead{(\textit{J}'\subs{K'}--\textit{J}\subs{K}} & \colhead{(GHz)} & \colhead{(K)} & \colhead{(K km s$^{-1}$)} & \colhead{(K km s$^{-1}$)}
}
\startdata
\multicolumn{5}{c}{Setting 1, UT 2017 April 11, \textit{r}\subs{H} = 1.14 au, $\Delta$ = 1.19 au} \\
\hline
$13_1$--$13_0 A^{-+}$ & 342.730 & 227.5 & 0.18 $\pm$ 0.04 & 0.025 $\pm$ 0.005\\
\hline
\multicolumn{5}{c}{Setting 2, UT 2017 April 15, \textit{r}\subs{H} = 1.12 au, $\Delta$ = 1.18 au} \\
\hline
$1_1$--$0_0$ $A^{+}$ & 350.905 & 16.8 & 0.40 $\pm$ 0.05 & 0.084 $\pm$ 0.017 \\
$4_0$--$3_{-1}$ $E$ & 350.687 & 36.3 & 0.34 $\pm$ 0.04 & 0.041 $\pm$ 0.008\\
$7_2$--$6_1$ $E$ & 363.739 & 87.3 & 0.35 $\pm$  0.10 & 0.061 $\pm$ 0.012
\enddata
\tablecomments{\sups{a} Spectrally integrated flux of spectra extracted from a nucleus-centered beam at the position of peak continuum flux. \sups{b} Spectrally integrated flux of autocorrelation spectra.
}
\end{deluxetable*}

\subsection{Expansion Velocities, Kinetic Temperature, and Asymmetric Outgassing}\label{subsec:vexp}
\begin{figure*}
\gridline{\fig{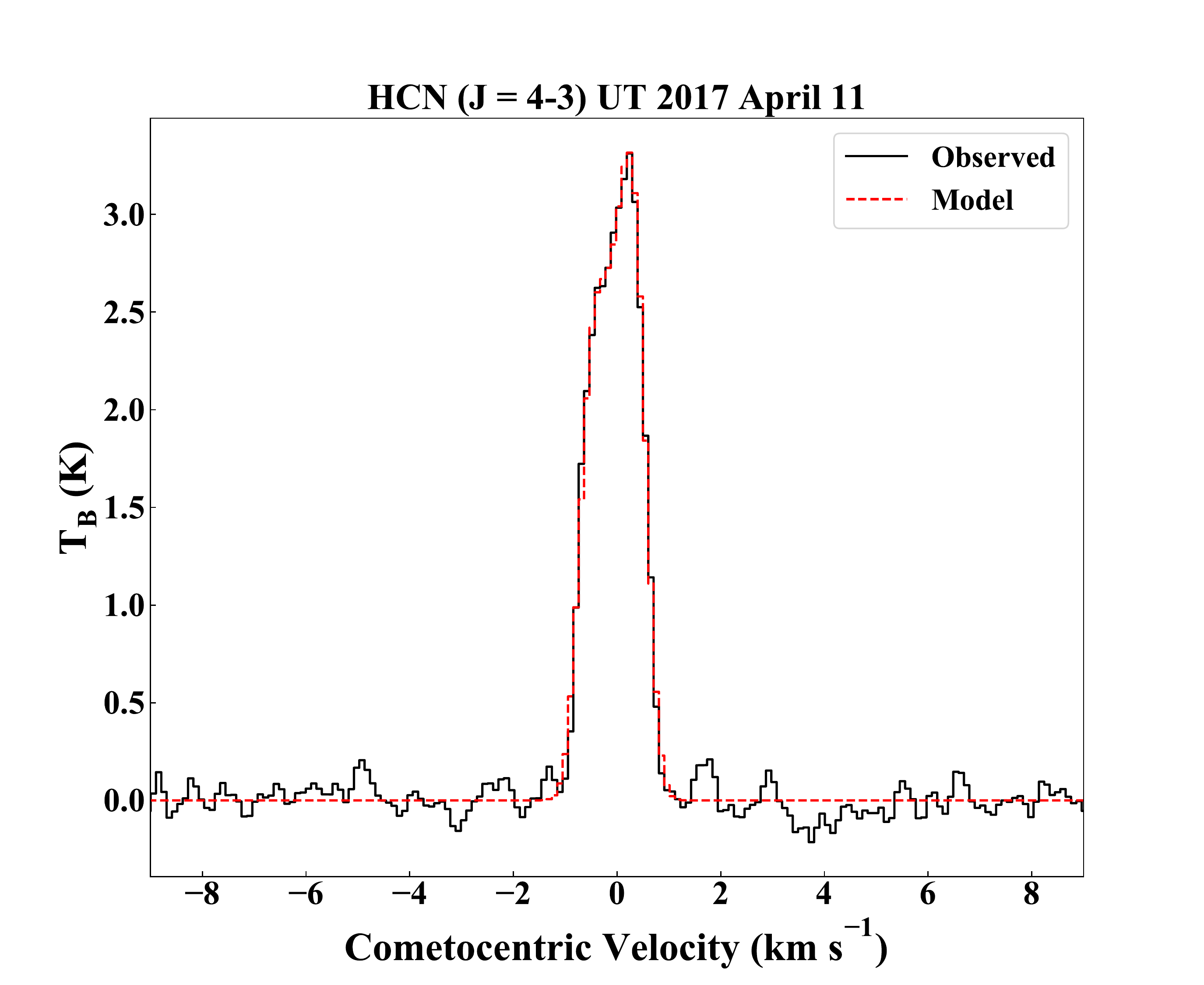}{0.48\textwidth}{(A)}
	}
\gridline{\fig{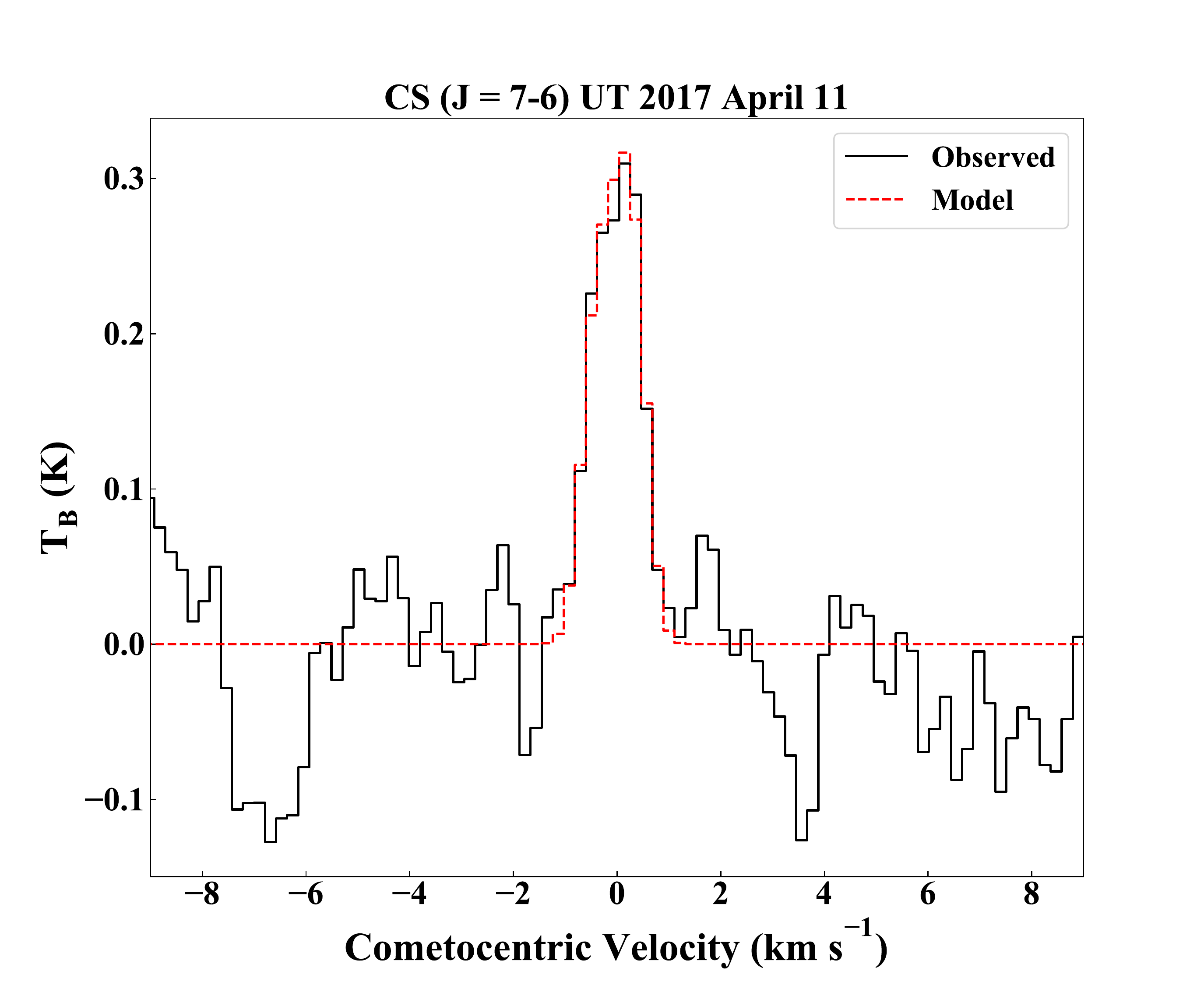}{0.48\textwidth}{(B)}
          \fig{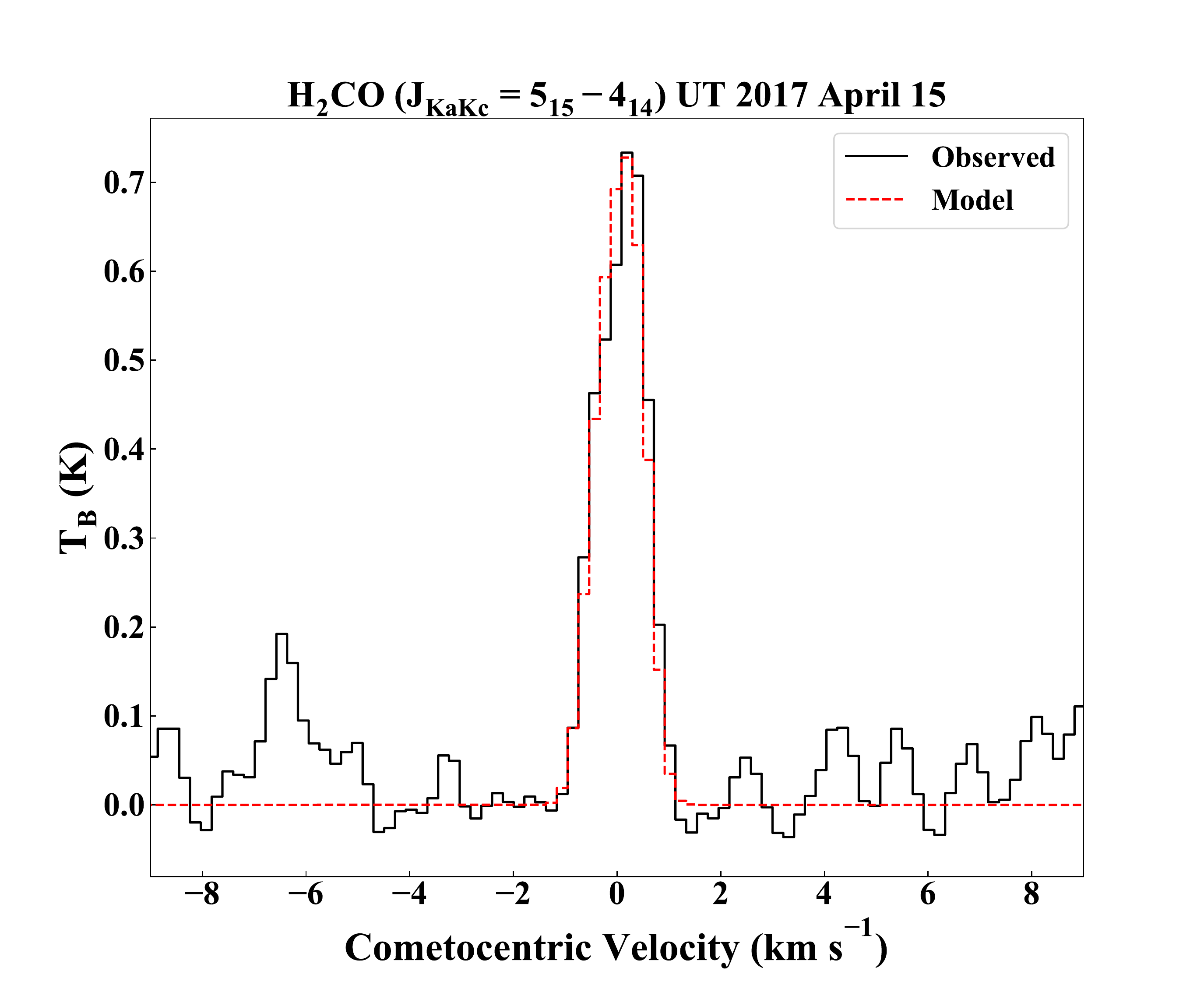}{0.48\textwidth}{(C)}
          }
\gridline{\fig{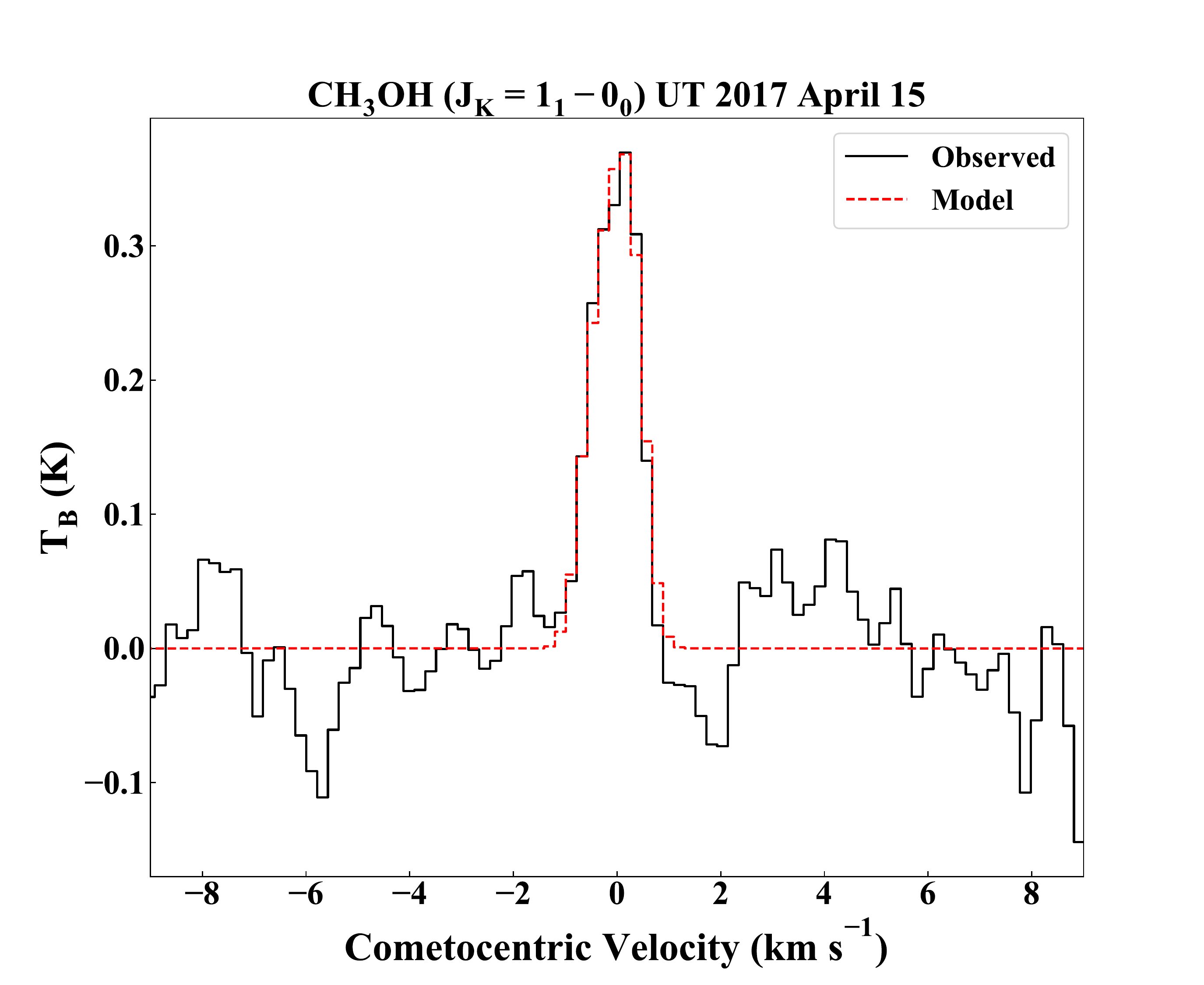}{0.48\textwidth}{(D)}
          \fig{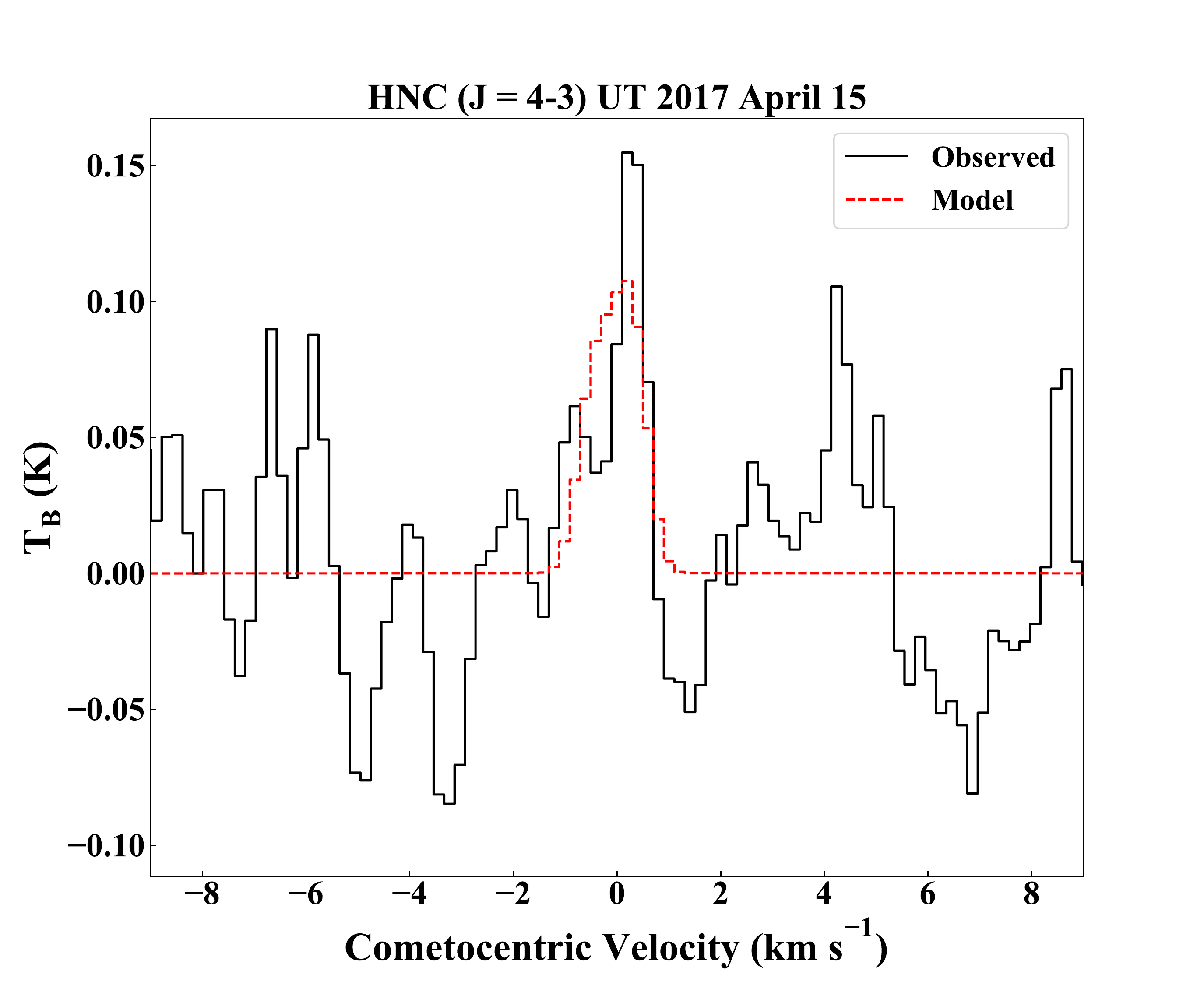}{0.48\textwidth}{(E)}
          }
\caption{\textbf{(A)--(E).} Extracted interferometric spectra and best-fit models for (A) HCN, (B) CS, (C) H\subs{2}CO, (D) CH\subs{3}OH, and (E) HNC. Spectra were extracted from the nucleus-centered beam.
\label{fig:spec}}
\end{figure*}

We derived expansion velocities, coma kinetic temperatures, and outgassing asymmetry factors for each molecule by performing nonlinear least-squares fits of our radiative transfer models to extracted interferometric and autocorrelation spectral line profiles. Table~\ref{tab:vexp} lists our results for the strongest line for each detected species and Figure~\ref{fig:spec} shows our extracted spectra and best-fit models for the interferometric spectra.

\subsubsection{Kinetic Temperature}\label{subsubsec:tkin}
\begin{figure}[ht!]
\plotone{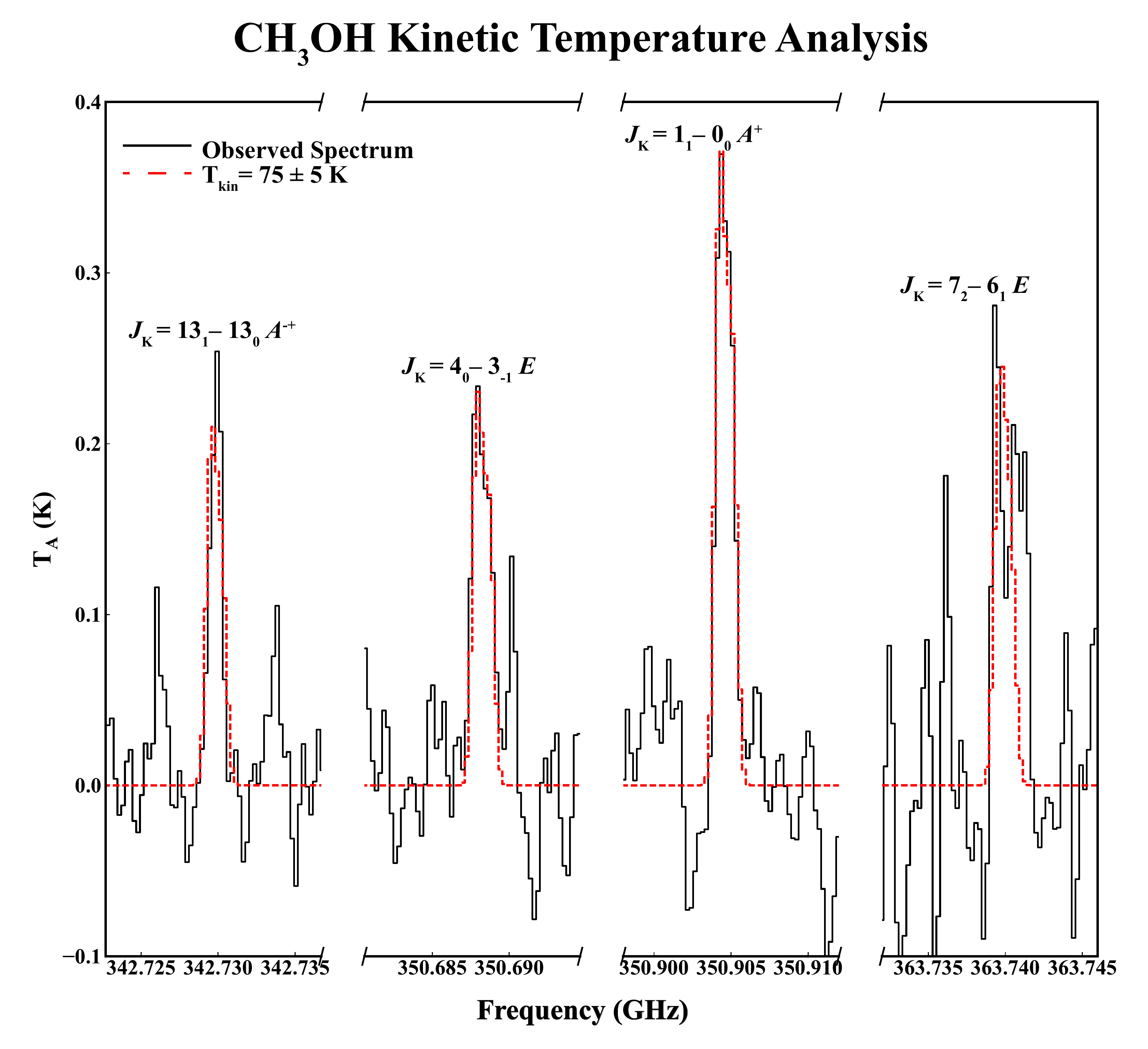}
\caption{Extracted interferometric spectra and best-fit kinetic temperature model (\textit{T}\subs{kin} = 75 $\pm$ 5 K) for the CH\subs{3}OH \textit{J}\subs{K} = $13_1$--$13_0$ $A^{-+}$ transition in ER61 on UT 2017 April 11 and the \textit{J}\subs{K} = $1_1$--$0_0$ $A^{+}$, \textit{J}\subs{K} = $4_0$--$3_{-1}$ $E$, and \textit{J}\subs{K} = $7_2$--$6_1$ $E$ transitions on UT 2017 April 15. Spectra were extracted from the nucleus-centered beam. \label{fig:tkin}}
\end{figure}

We determined the coma kinetic temperature by fitting full non-LTE models to all four detected CH\subs{3}OH lines simultaneously for both the interferometric and autocorrelation spectra. We assumed that the collision rate between CH$_3$OH and H$_2$O is the same as that with H$_2$ \citep[taken from the LAMDA database;][]{Schoier2005,Rabli2010}. Calculating formal CH$_3$OH--H$_2$O collision rates is the subject of a future work and beyond the scope of this paper. For the interferometric nucleus-centered spectra, the best-fit kinetic temperature, \textit{T}\subs{kin} = 75 $\pm$ 5 K, is higher than the derived rotational temperature (61 $\pm$ 5 K). Similarly, for the autocorrelation spectra, the best-fit kinetic temperature is \textit{T}\subs{kin} = 88 $\pm$ 8 K, and is significantly higher than the rotational temperature (58 $\pm$ 4 K). As these two measurements are in formal agreement, we assumed a kinetic temperature \textit{T}\subs{kin} = 78 K (the weighted average of the best-fit kinetic temperatures) when modeling all other molecules in ER61. Production rates derived by setting \textit{T}\subs{kin} = \textit{T}\subs{rot} (61 K) are in formal agreement with those derived by setting \textit{T}\subs{kin} = 78 K. Figure~\ref{fig:tkin} shows extracted spectra and our best-fit model for each CH\subs{3}OH line.

\subsubsection{Expansion Velocities and Asymmetric Outgassing}\label{subsubsec:asymmetric}
Uniform outgassing vs. hemispheric asymmetry along the Sun--comet vector were considered when fitting the observations for each molecule, beginning with the HCN (\textit{J} = 4 -- 3) transition, the strongest in our dataset. Assuming isotropic outgassing, our best-fit radiative transfer model finds \textit{v}\subs{exp} = 0.51 $\pm$ 0.02 km s\sups{-1}, with a reduced chi-square ($\chi^2_r$) of 2.61, corresponding to \textit{P} = 0.0 (where \textit{P} is the probability that the model differs from the observations due to random noise). Instead, assuming hemispheric asymmetry along the Sun--comet vector, our best-fitting model indicates an expansion velocity \textit{v}\subs{exp}(1) = 0.60 $\pm$ 0.01 km s\sups{-1} in the sunward hemisphere and \textit{v}\subs{exp}(2) = 0.42 $\pm$ 0.01 km s\sups{-1} in the anti-sunward hemisphere, with a ratio of sunward to anti-sunward production rates of 1.30 $\pm$ 0.05. For the asymmetric model, we find $\chi^2_r$  = 1.005, corresponding to \textit{P} = 0.47; therefore, the asymmetric outgassing model is strongly favored. 

Expansion velocities and asymmetry factors for CS, CH\subs{3}OH, and HNC are consistent with those for HCN within uncertainty for the interferometric spectra, suggesting similar outgassing for each species and/or their parents. The exception was H\subs{2}CO, where the best-fitting asymmetry factor of sunward to anti-sunward production rates was 0.51 $\pm$ 0.09 (Table~\ref{tab:vexp}). This suggests increased H\subs{2}CO production in the anti-sunward hemisphere of ER61.

The slower expansion velocity in the anti-sunward hemisphere combined with the asymmetry factor (1.30), indicating higher HCN production in the sunward hemisphere, helps to explain the red-shifted line profile, as HCN molecules outgassing from the anti-sunward side would have remained in the ACA beam for longer than those from the sunward side. Conversely, for H$_2$CO, the expansion velocities in each hemisphere are in formal agreement, yet the asymmetry factor (0.51) indicates increased H$_2$CO production in the anti-sunward hemisphere. This helps explain how the distinct H$_2$CO outgassing pattern (compared to all other species) still results in a red-shifted line profile -- H$_2$CO molecules remained in the ACA beam for roughly the same time in each hemisphere, yet there was a greater amount of H$_2$CO production in the anti-sunward (red) hemisphere.

Compared with quantities derived for the interferometric spectra, the expansion velocities derived from the autocorrelation spectra were in general slightly larger, with the HCN expansion velocities being most similar in both cases and the H$_2$CO, CH$_3$OH, and CS expansion velocities showing significant differences. Asymmetry factors (the ratio of sunward to anti-sunward production) were similar for all molecules except CS, for which the asymmetry factor is a factor of two higher for the interferometric spectra compared to the autocorrelation spectra, suggesting increased CS production in the anti-sunward hemisphere of ER61 on large angular scales. Combined with the non-uniform coma structures displayed in the channel maps of each molecule (Figure~\ref{fig:chanmaps}), these results highlight the complex nature of ER61’s coma.

\begin{deluxetable*}{ccccccc}
\tablenum{4}
\tablecaption{Expansion Velocities, Kinetic Temperature, and Asymmetry Factors in C/2015 ER61 (PanSTARRS)\label{tab:vexp}}
\tablewidth{0pt}
\tablehead{
\colhead{Molecule} & \colhead{\textit{T}\subs{kin}\sups{a}} & \colhead{\textit{v}\subs{exp}(1)\sups{b}} & \colhead{\textit{v}\subs{exp}(2)\sups{c}} & \colhead{\textit{Q}\subs{1}/\textit{Q}\subs{2}\sups{d}} & \colhead{$\chi^2$\subs{r}\sups{e}} & \colhead{\textit{P}\sups{f}} \\
\colhead{(Transition)} & \colhead{(K)} & \colhead{(km s\sups{-1})} & \colhead{(km s\sups{-1})} & \colhead{} & \colhead{} & \colhead{}
}
\startdata
\multicolumn{7}{c}{Interferometric Spectra} \\
\hline
HCN (4--3) & (78) & 0.60 $\pm$ 0.01 & 0.42 $\pm$ 0.01 & 1.30 $\pm$ 0.05 & 1.001 & 0.46 \\
                   &        & \multicolumn{2}{c}{0.51 $\pm$ 0.02\sups{g}} & $\cdot \cdot \cdot$ & 2.610\sups{g} & 0.00\sups{g} \\
CS (7--6) & (78) & 0.55 $\pm$ 0.08 & 0.41 $\pm$ 0.06 & 1.2 $\pm$ 0.4 & 0.907 & 0.78 \\
H\subs{2}CO ($5_{15}$--$4_{14}$) & (78) & 0.54 $\pm$ 0.04 & 0.50 $\pm$ 0.02 & 0.51 $\pm$ 0.09 & 1.073 & 0.25 \\
                                                         & (78) & (0.60) & (0.42) & (1.30) & 1.518 & 0.00\sups{h} \\
CH\subs{3}OH ($1_1$--$0_0$ $A^{+}$) & 75 $\pm$ 5 & 0.56 $\pm$ 0.07 & 0.37 $\pm$ 0.06 & 1.2 $\pm$ 0.4 & 1.043 & 0.34 \\
HNC (4--3) & (78) & 0.9 $\pm$ 0.3 & 0.4 $\pm$ 0.1 & 1.1 $\pm$ 0.6 & 0.948 & 0.65 \\
                  & (78) & (0.60) & (0.42) & (1.30) & 0.967 & 0.60 \\ 
\hline
\multicolumn{7}{c}{Autocorrelation Spectra} \\
\hline
HCN (4--3) & (78) & 0.65 $\pm$ 0.01 & 0.50 $\pm$ 0.01 & 1.19 $\pm$ 0.03 & 1.377 & 0.001 \\
CS (7--6) & (78) & 0.68 $\pm$ 0.05 & 0.60 $\pm$ 0.03 & 0.61 $\pm$ 0.07 & 1.300 & 0.033 \\
H\subs{2}CO ($5_{15}$--$4_{14}$) & (78) & 0.84 $\pm$ 0.08 & 0.48 $\pm$ 0.04 & 0.7 $\pm$ 0.1 & 1.2988 & 0.032 \\
CH\subs{3}OH ($1_1$--$0_0$ $A^{+}$) & 88 $\pm$ 8 & 0.76 $\pm$ 0.02 & 0.63 $\pm$ 0.03 & 1.7 $\pm$ 0.2 & 1.306 & 0.002 \\
HNC (4--3) & (78) & 0.9 $\pm$ 0.2 & 0.6 $\pm$ 0.1 & 1.4 $\pm$ 0.5 & 1.064 & 0.317 \\
\enddata
\tablecomments{\sups{a} Coma kinetic temperature. We assumed the weighted average of the kinetic temperatures measured from CH$_3$OH interferometric and autocorrelation spectra when modeling other species. Values in parentheses are assumed. \sups{b} Expansion velocity in the sunward hemisphere. \sups{c} Expansion velocity in the anti-sunward hemisphere. \sups{d} Asymmetry factor: ratio of production rates in the sunward vs. anti-sunward hemisphere. \sups{e} Reduced $\chi^2$ of the model fit. \sups{f} \textit{P}-value of the fit: the probability that the difference between the model and the observations is due to random noise. \sups{g} Expansion velocity, reduced $\chi^2$, and \textit{P}-value for the best-fit HCN model assuming isotropic outgassing. \sups{h} Reduced $\chi^2$ and \textit{P}-value for H\subs{2}CO when assuming the expansion velocities and asymmetry factor found for HCN, indicating that increased anti-sunward production is statistically strongly favored for H\subs{2}CO.
}
\end{deluxetable*}

\subsection{CO Autocorrelation Spectra}\label{subsec:CO}
Emission from the CO (\textit{J} = 3 -- 2) transition near 345.795 GHz was not confirmed in our interferometric maps. To constrain CO production in ER61, we extracted a spectrum from the expected position of the nucleus in the CO map. CO was assumed to be outgassing with the same expansion velocities and asymmetry factor as HCN. Integrating from -1.5 km s\sups{-1} to +1.5 km s\sups{-1} (the width of the HCN (\textit{J} = 4 -- 3) line), a 3$\sigma$ upper limit for the integrated intensity of $<$ 147 mK km s\sups{-1} was derived, corresponding to a production rate \textit{Q}(CO) $<$ 2.3 $\times$ 10\sups{27} mol s\sups{-1} and a mixing ratio CO/H\subs{2}O $<$ 1.96\%.

In order to improve sensitivity and derive a more stringent upper limit for CO, we extracted autocorrelation spectra as in Section~\ref{subsec:autocorr}. CO was not detected in the autocorrelation spectra. Based on the flux at the position of the CO (\textit{J} = 3 -- 2) transition in the autocorrelation spectra, we constrained a 3$\sigma$ upper limit for the integrated intensity of $<$ 9 mK km s\sups{-1}, corresponding to a production rate \textit{Q}(CO) $<$ 1.91 x 10\sups{27} mol s\sups{-1} and a mixing ratio CO/H\subs{2}O $<$ 1.59\%. CO mixing ratios span a wide range in measured Oort cloud comets (0.4 -- 23\%) and much narrower range in the fewer Jupiter-family comets for which this molecule has been detected (0.3 -- 4.3\%), making the upper limit in ER61 consistent with depleted Oort cloud comets and more similar to Jupiter-family comets \citep{DelloRusso2016a,Bockelee-Morvan2017}. This may suggest that ER61 formed in an area of the protosolar disk where ices suffered enhanced processing before incorporation into its nucleus compared to those from CO-rich Oort cloud comets. 

%On the other hand, the CO-depletion may be a reflection of its dynamical history: with a period \textit{P} $\sim$ 8600 years, the 2017 apparition was not ER61's first passage into the inner solar system. It is possible that previous, repeated perihelion passages depleted ER61's most volatile ices, such as CO.

\section{Production Rates and Molecular Distribution}\label{sec:prod}
We calculated molecular production rates and parent scale lengths for all detected species. Although our measurements did not sample H$_2$O, we calculated mixing ratios to facilitate comparisons with other comets. We adopted \textit{Q}(H$_2$O) = 1.2 $\times$ $10^{29}$ s$^{-1}$ based on near-infrared measurements of H$_2$O on UT 2017 April 15, coinciding with the second date of our ACA study \citep{Saki2021}. Following the methods of \cite{Boissier2007}, we performed nonlinear least-squares fits of models generated over a range of production rates and parent scale lengths to the real part of the interferometric and autocorrelation visibilities as a function of projected baseline for each molecule. Using a Haser model \citep{Haser1957}, we determined the distribution of each trace species using:
\begin{equation}
    n_d(r) = \frac{Q}{4 \pi v_{exp} r^2}\frac{\frac{v_{exp}}{\beta_d}}{\frac{v_{exp}}{\beta_d}-L_p}\left[\exp{\left(-\frac{r\beta_d}{v_{exp}}\right)}-\exp{\left(-\frac{r}{L_p}\right)}\right],
\end{equation}
\noindent where \textit{Q} is the production rate (molecules s$^{-1}$), \textit{v}\subs{exp} is the expansion velocity (km s$^{-1}$), \textit{$\beta$}\subs{d} is the photodissociation rate (s$^{-1}$), and \textit{L}\subs{p} is the parent scale length (km), with \textit{Q}, \textit{v}\subs{exp}, and \textit{L}\subs{p} allowed to vary as free parameters. 

\begin{figure*}
\gridline{\fig{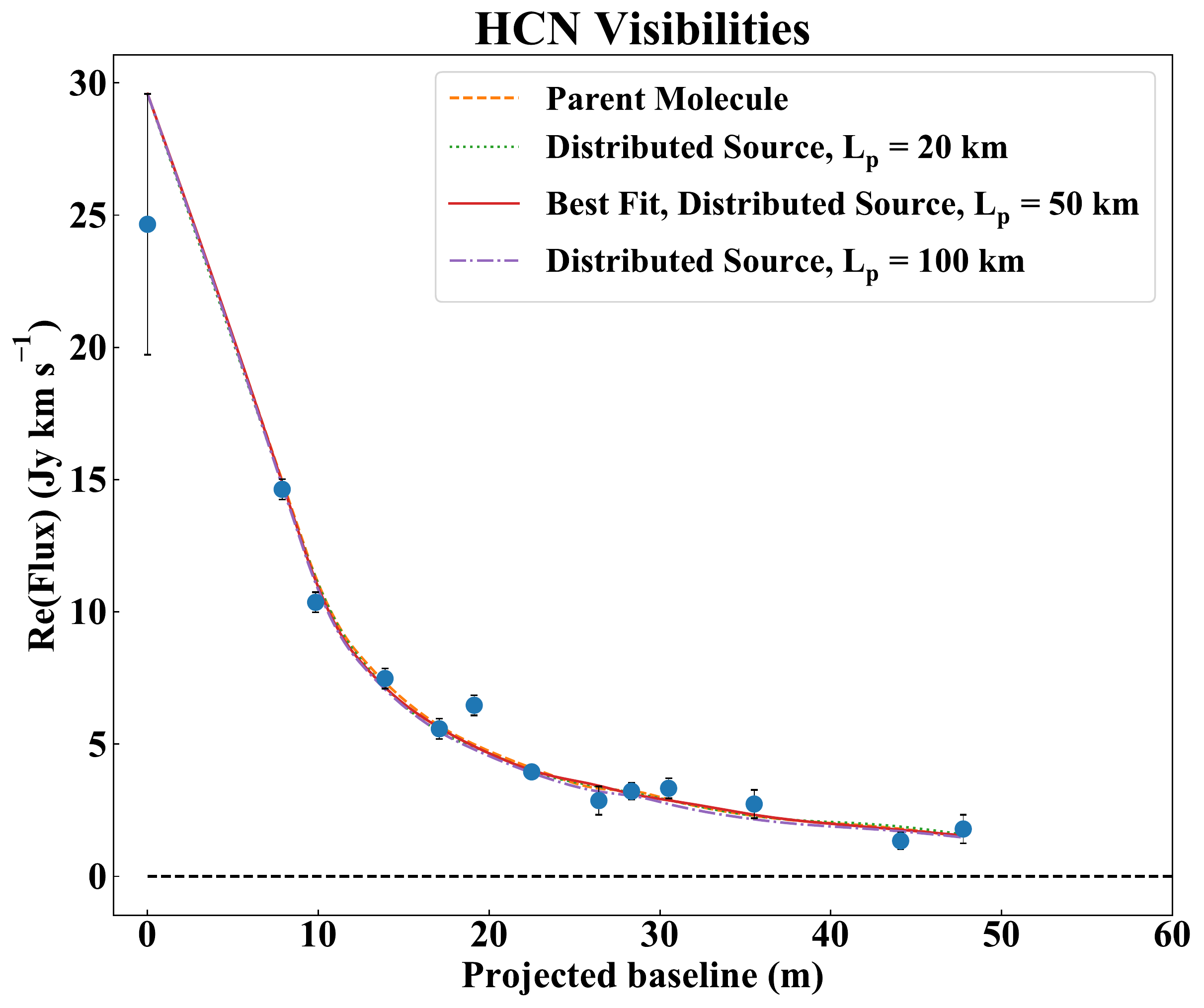}{0.45\textwidth}{(A)}
              \fig{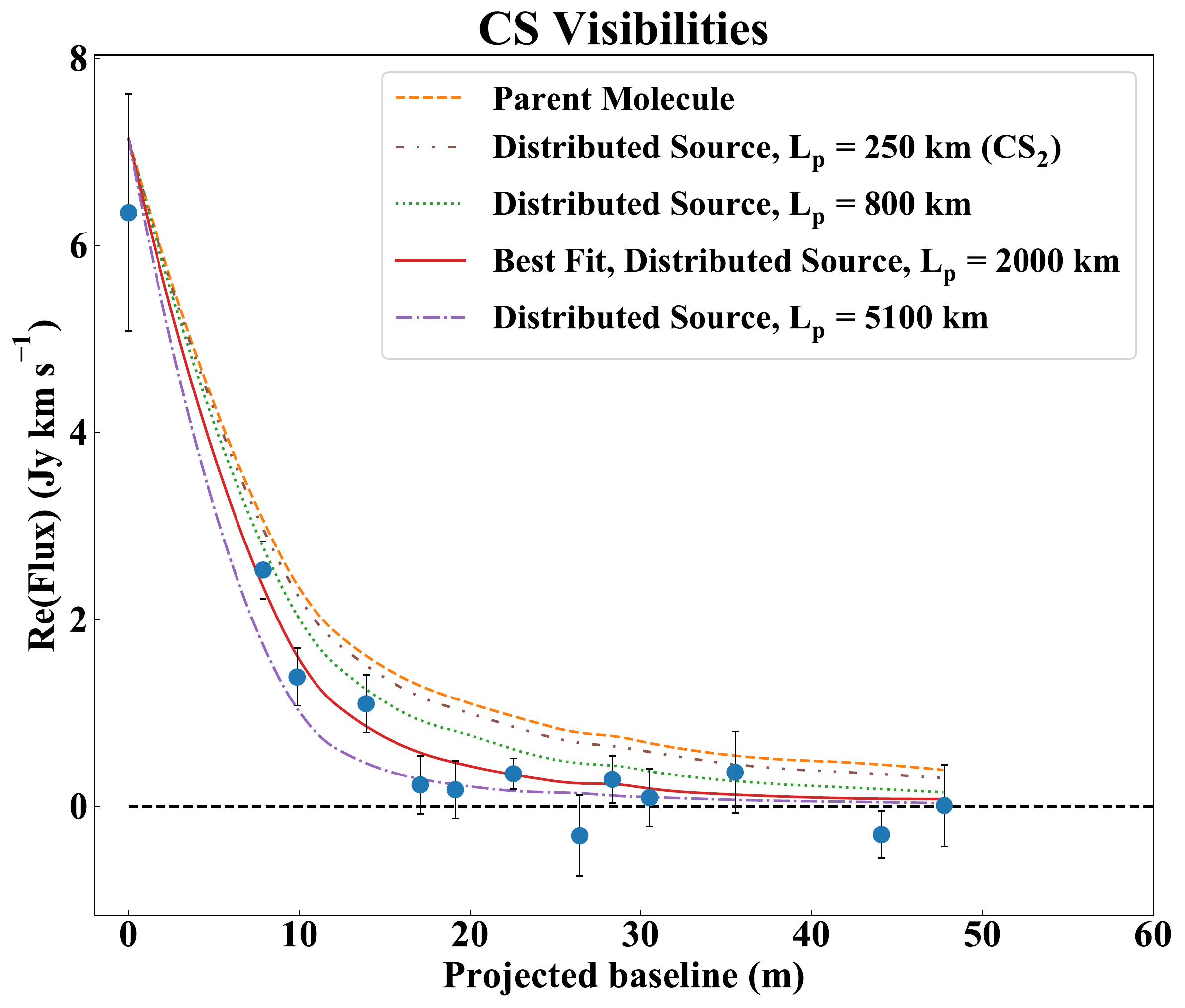}{0.45\textwidth}{(B)}
	}
\caption{\textbf{(A)--(B).} Real part of the observed visibility amplitude vs. projected baseline length for HCN and CS in ER61 on UT 2017 April 11. Modeled visibility curves for various parent scale lengths (\textit{L}\subs{p}) are shown with the best-fit model for each molecule plotted in red (solid) and $\pm$1$\sigma$ uncertainties in \textit{L}\subs{p} shown in green (dotted) and purple (dash--dotted). A parent molecule model (\textit{L}\subs{p} = 0 km, orange, dashed) is also shown for comparison. Interpolation is used between the zero-point spacing and shortest ACA baseline.
\label{fig:visplots1}}
\end{figure*}

\begin{figure*}
\gridline{\fig{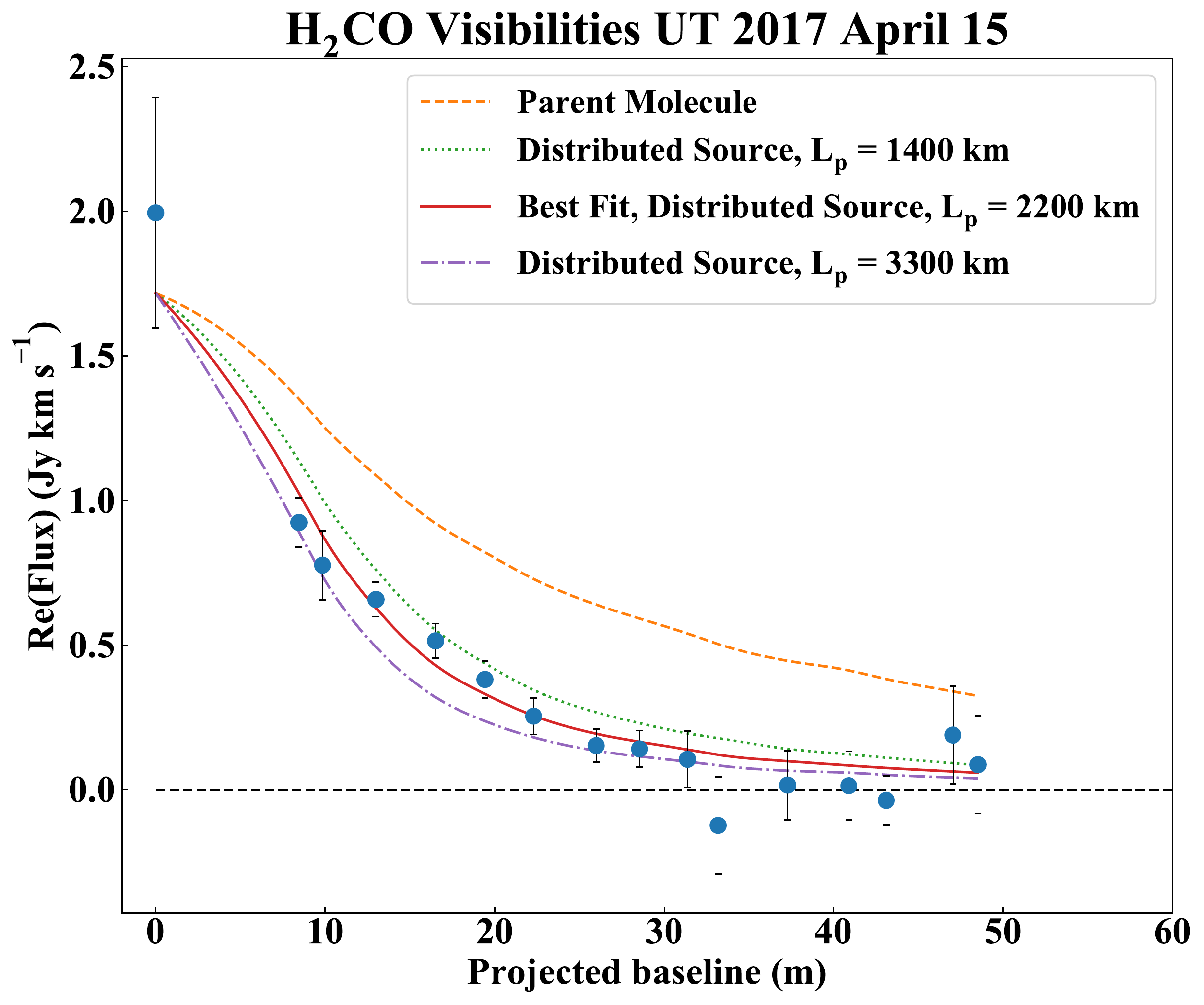}{0.45\textwidth}{(A)}
          \fig{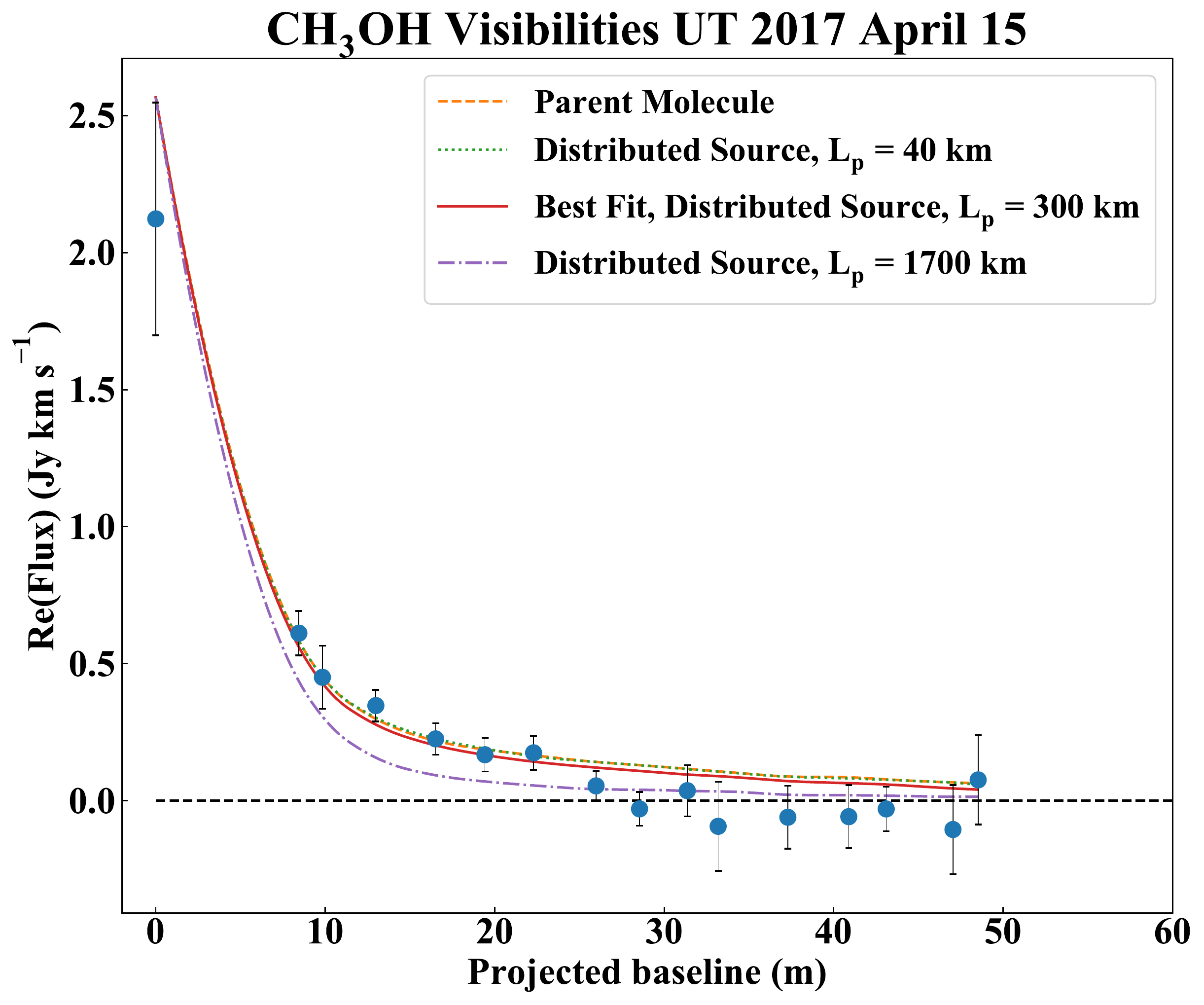}{0.45\textwidth}{(B)}
          }
\gridline{\fig{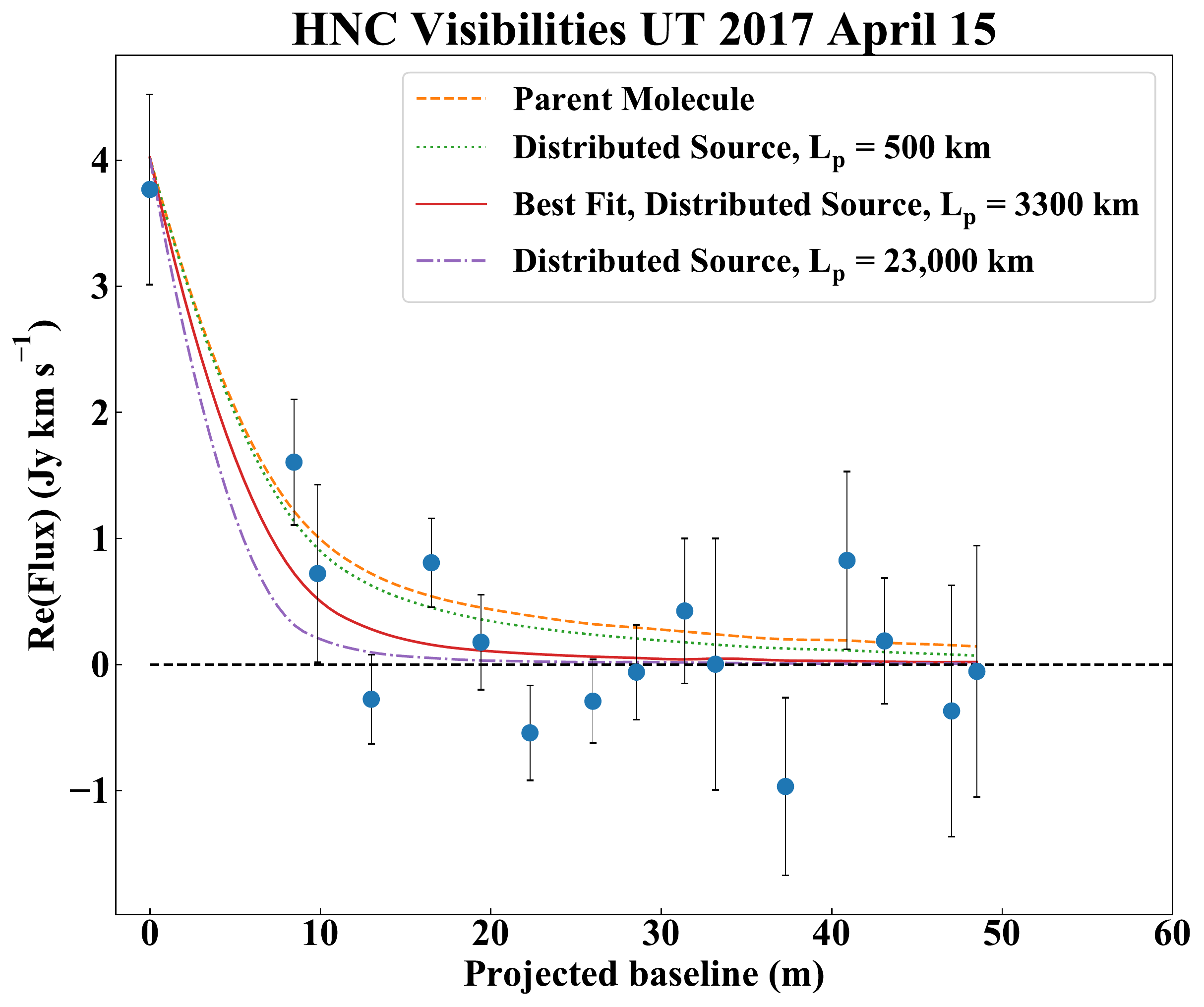}{0.45\textwidth}{(C)}
         }
\caption{\textbf{(A)--(C).} Real part of the observed visibility amplitude vs. projected baseline length for H\subs{2}CO, CH\subs{3}OH, and HNC in ER61 on UT 2017 April 15, with traces and labels as in Figure~\ref{fig:visplots1}. Interpolation is used between the zero-point spacing and shortest ACA baseline.
\label{fig:visplots2}}
\end{figure*}

We used the expansion velocities, asymmetry factors, and kinetic temperatures listed in Table~\ref{tab:vexp} for each molecule when generating synthetic visibilities for each component of the model (interferometric and autocorrelation). A single \textit{L}\subs{p} and \textit{Q} was used to fit both components (interferometric and autocorrelation). Figures~\ref{fig:visplots1} and~\ref{fig:visplots2} show the observed visibilities with best-fit models, enabling us to discern parent species from product and to determine the mechanism of molecular production for each species in ER61. We detail our results for each molecule in turn.

\subsection{HCN}\label{subsec:hcn}
We detected emission from the HCN (\textit{J} = 4 -- 3) transition near 354.505 GHz. We find a best-fit parent scale length \textit{L}\subs{p} = $50^{+50}_{-30}$ km, consistent with HCN production at or very close to the nucleus (Figure~\ref{fig:visplots1}). This is consistent with results from previous ALMA studies using the 12 m array \citep{Cordiner2014} indicating HCN production as a parent species, although our ACA measurements lack the longer baselines necessary to explicitly confirm HCN release at the nucleus or in the very inner coma. The production rate, \textit{Q}(HCN) = (8.6 $\pm$ 0.3) $\times$ $10^{25}$ s$^{-1}$, corresponds to an abundance relative to H$_2$O of 0.072\%, a value in the low range of HCN abundances measured in comets at radio wavelengths \citep{Bockelee-Morvan2017}.

\subsection{CS}\label{subsec:cs}
Previous work for other comets has indicated that CS is a product species produced in the inner coma, ranging in abundance from 0.02 -- 0.20\% with respect to H\subs{2}O. Although short-lived CS$_2$ has been suggested, a precursor has never been conclusively identified \citep{Feldman2004,Boissier2007,Bogelund2017}. Our visibility modeling is well aligned with CS as a product species, with parent scale length \textit{L}\subs{p} = $2000^{+3100}_{-1200}$ km, corresponding to a photodissociation rate $\beta$ = $(3.57^{+5.36}_{-2.17})$ $\times$ $10^{-4}$ s$^{-1}$ at \textit{r}\subs{H} = 1 au. This is considerably longer than the photodissociation rate expected for CS$_2$  (2.93 $\times$ $10^{-3}$ s$^{-1}$ at \textit{r}\subs{H} = 1 au), and suggests the presence of an unidentified parent for CS production in ER61. Although CS production from CS$_2$ in the very inner coma cannot be ruled out by our measurements, Figure~\ref{fig:visplots1} demonstrates that CS$_2$ photolysis cannot account for the significant extended production of CS revealed in ER61. The production rate, \textit{Q}(CS) = (7.4 $\pm$ 0.9) $\times$ $10^{25}$ s$^{-1}$, corresponds to an abundance relative to H$_2$O of 0.061\%, consistent with average abundances in comets \citep{Bockelee-Morvan2017}.

The presence of a distributed source of CS in the coma would not be not surprising in the context of other species found in comets (e.g., H$_2$CO, HNC) as well as unresolved questions regarding the sulfur inventory of various astronomical sources. An observed depletion of gas-phase sulfur-bearing species in dense clouds and star-forming regions along with a depletion of ice-phase sulfur-bearing molecules in the interstellar medium has resulted in a search for ``missing sulfur'' \citep{Mifsud2021}. Similar to the unknown refractory parent suggested for H$_2$CO (Section~\ref{subsec:h2co}) and the refractory ammonium salts identified as potential reservoirs for the ``missing nitrogen'' in comets \citep{Altwegg2020}, an unidentified refractory sulfur-bearing component has been suggested to address the observed sulfur depletion. A distributed, refractory source for CS would be consistent with the observed heliocentric dependence of CS production \citep[][and references therein]{Cottin2008}, which cannot be explained by photolysis of CS$_2$, as well as the organo-sulfur molecules identified alongside the ammonium salts in comet 67P/Churyumov--Gerasimenko by Rosetta \citep{Altwegg2020}. Thus, it is possible that an unidentified refractory sulfur-bearing parent may be responsible for the extended CS production in ER61, analogous to those suspected for H$_2$CO and HNC. 

On the other hand, the photolysis of gas-phase species provides another potential avenue for extended CS production. \cite{Jackson1986} examined potential parents of CS in comet IRAS-Araki-Alcock 1983d, for which the CS parent scale length was $\sim$300 km. HNCS (thiocyanic acid) was discounted as only producing the HNC, H, S, and NCS fragments. CH$_3$SH (methanethiol) would require cleaving four H's which would likely result in poor yields. Furthermore, CH$_3$SH dissociates to CH$_3$S + H and CH$_3$ + SH rapidly \citep[calculated as $\sim$500 s at \textit{r}\subs{H} = 1 au;][]{Huebner2015}. Perhaps the most compatible gas-phase precursor previously detected in comets is H$_2$CS (thioformaldehyde). Ranging from 0.009 -- 0.09\% \citep{Bockelee-Morvan2017} in measured comets, H$_2$CS was associated with coma production as well as direct nucleus release in comet 67P/Churyumov--Gerasimenko by Rosetta \citep{Calmonte2016}. Although its lifetime in the solar radiation field has not been published, it may be similar to H$_2$CO \citep{Jackson1986}, suggesting a lifetime of $\sim$5600 s at the heliocentric distance of ER61 and a scale length of $\sim$3400 km, in good agreement with our derived \textit{L}\subs{p} of 2000$^{+3100}_{-1200}$ km for CS. Although the observed abundance range of H$_2$CS in comets encompasses that for CS in ER61 reported here, it is too low to account for comets with higher CS abundances (up to 0.2\%). 

Clearly, the identity of the parent for extended CS production in ER61 is unresolved. The lack of strong correlation with a specific parent species highlights the need for laboratory measurements of potential CS parents.  This includes studies of not only the photodissociation rates but of products and branching ratios for incorporation into analytical and chemical models.

\subsection{H\subs{2}CO}\label{subsec:h2co}
 We detected emission from the H\subs{2}CO (\textit{J}\subs{KaKc} = $5_{15}$ -- $4_{14}$) transition near 351.768 GHz. Although H\subs{2}CO has been primarily associated with nucleus sources based on observations of the inner coma at near-infrared wavelengths \citep{DelloRusso2016}, measurements from the Neutral Mass Spectrometer on board the Giotto spacecraft indicated distributed source production of H\subs{2}CO in comet 1P/Halley \citep{Meier1993}. Furthermore, millimeter studies have indicated H\subs{2}CO production from an unknown parent source with a scale length of 6800 km at 1 au \citep{Biver1999,Bockelee2000,Milam2006}, and ALMA observations of the inner coma suggest a scale length of 1000 -- 5000 km \citep{Cordiner2014,Cordiner2017b}. The H\subs{2}CO visibilities in ER61 are similarly consistent with distributed source production, with a best-fit parent scale length \textit{L}\subs{p} = $2200^{+1100}_{-800}$ km at \textit{r}\subs{H} $\sim$ 1.12 au. Our production rate, \textit{Q}(H\subs{2}CO) = (3.7 $\pm$ 0.3) $\times$ 10\sups{26} s\sups{-1}, corresponds to an H\subs{2}CO mixing ratio of 0.31\%, consistent with average values in comets \citep{DelloRusso2016a,Bockelee-Morvan2017}. 

Our measured \textit{L}\subs{p} is consistent with the parent scale length \textit{L}\subs{p} = $1200^{+1200}_{-400}$ km found in C/2012 F6 (Lemmon) at \textit{r}\subs{H} = 1.48 au using the 12 m array, but considerably larger than the 280 $\pm$ 50 km parent scale length found in C/2012 S1 (ISON) at \textit{r}\subs{H} = 0.54 au \citep{Cordiner2014}. Taking into account the greatly increased solar radiation at the small \textit{r}\subs{H} of S1 (ISON) compared to ER61 and F6 (Lemmon), these differing parent scale lengths are all compatible with optically thin photodissociation of a parent molecule in uniform outflow.

On the other hand, a comparison of the derived photodissociation rates at \textit{r}\subs{H} = 1 au for all three comets (Table~\ref{tab:beta}) indicates a considerably longer photodissociation scale for H$_2$CO in ER61 than for S1 (ISON) or F6 (Lemmon). We derived these ``effective'' parent photodissociation rates at \textit{r}\subs{H} = 1 au by correcting \textit{L}\subs{p} for the gas expansion velocity (\textit{v}\subs{exp}) reported for each comet and scaling by the \textit{r}\subs{H} of the comet as \textit{r}\subs{H}$^{2}$. Assuming \textit{v}\subs{exp} = 0.8 km s$^{-1}$ at \textit{r}\subs{H} = 1 au, the parent scale length of 6800 km mentioned above would correspond to $\beta$ = 1.17 $\times$ $10^{-4}$ s$^{-1}$, in better agreement with the value derived for ER61 than that for S1 (ISON) or F6 (Lemmon). The differences seen among comets may indicate non-Haser behavior for the unidentified H$_2$CO parent, and emphasize the work remaining in understanding the H$_2$CO photochemistry in comets. 

\begin{deluxetable*}{cccc}
\tablenum{5}
\tablecaption{Parent Photodissociaton Rates at \textit{r}\subs{H} = 1 au in Comets Studied with ALMA \label{tab:beta}}
\tablewidth{0pt}
\tablehead{
\colhead{Molecule} & \colhead{ER61} & \colhead{C/2012 S1 (ISON)} & \colhead{C/2012 F6 (Lemmon)} 
}
\startdata
HCN & ($1.55^{+2.34}_{-0.78}$) $\times$ $10^{-2}$& $>5.8$ $\times$ $10^{-3}$ & $>3.05$ $\times$ $10^{-2}$ \\
CS & ($3.57^{+5.36}_{-2.17})$ $\times$ $10^{-4}$ & ($1.46^{+0.47}_{-0.29})$ $\times$ $10^{-3}$ & $\cdot \cdot \cdot$ \\
H$_2$CO & ($3.08^{+1.76}_{-1.03})$ $\times$ $10^{-4}$ & ($1.04^{+0.23}_{-0.16})$ $\times$ $10^{-3}$ & ($1.26^{+0.63}_{-0.63})$ $\times$ $10^{-3}$\\
HNC & ($2.28^{+12.7}_{-1.95})$ $\times$ 10$^{-4}$ & ($4.16^{+5.55}_{-2.54})$ $\times$ $10^{-4}$ & $\cdot \cdot \cdot$
\enddata
\tablecomments{Photodissociation rate (s$^{-1}$) corrected for \textit{v}\subs{exp} and \textit{r}\subs{H} at the time of the observations. Rates for ISON and Lemmon taken from \cite{Cordiner2014} except for CS \citep{Bogelund2017}.
}
\end{deluxetable*}

\subsection{CH\subs{3}OH} \label{subsec:ch3oh}
As detailed in Section~\ref{subsec:trot}, we detected multiple transitions of CH$_3$OH in ER61. The visibilities are best fit with a parent scale length \textit{L}\subs{p} = $300^{+1400}_{-260}$ km. The uncertainty in the production rate, \textit{Q}(CH$_3$OH) = (2.3 $\pm$ 0.9) $\times$ $10^{27}$ s$^{-1}$, reflects the significant uncertainty in \textit{L}\subs{p} rather than the spectral line (detected at 8$\sigma$; Table~\ref{tab:comp}), and corresponds to a mixing ratio of $\sim$1.9\%.

The half-width at half-maximum of the ACA beam on April 15 corresponded to nucleocentric distances of 988 km $\times$ 2092 km, so it is possible that an unresolved distributed source of CH$_3$OH production was captured by our measurements. Despite its high nominal value, it is important to note that the lower bound on \textit{L}\subs{p} is consistent with CH$_3$OH production at or very near to the nucleus, and our measurements cannot decisively distinguish between direct nucleus release and distributed source production of CH$_3$OH in ER61.

If CH$_3$OH was indeed produced from distributed sources in ER61, it may have been from icy CH$_3$OH-coated grains. Icy grain release of CH$_3$OH was indicated for comet 103P/Hartley 2 \citep{Drahus2012,Boissier2014} as well as for comet 46P/Wirtanen \citep{Bonev2021,Roth2021a}, although these were both ``hyperactive'' comets displaying significant production of H$_2$O from icy grain release \citep[e.g.,][and references therein]{Lis2019}. Although ER61 was not a hyperactive comet, as noted in Section~\ref{sec:cont}, it had undergone a significant outburst 11.5 days prior and was likely in the process of nucleus fragmentation during our observations \citep{Sekanina2017}. This nucleus fragmentation process may have liberated icy CH$_3$OH-coated grains into the coma, from which CH$_3$OH sublimed and added to the overall production rate. 

\subsection{HNC}\label{subsec:hnc}
Finally, we detected emission from the HNC (\textit{J} = 4 -- 3) transition near 362.630 GHz. Early studies of HNC in C/1995 O1 (Hale--Bopp) were inconsistent with nucleus production \citep{Irvine1998}, and recent work with ALMA in C/2012 S1 (ISON) suggests that HNC is produced via degradation of complex organics, perhaps HCN polymers \citep{Cordiner2017b}. Our best-fit parent scale length is \textit{L}\subs{p} = $3300^{+19,700}_{-2800}$ km, which is consistent with a dominant coma source for this molecule. We derive a production rate \textit{Q}(HNC) = (8.2 $\pm$ 1.7) $\times$ 10\sups{24} s\sups{-1}, which corresponds to an HNC mixing ratio of 0.007\% with respect to H\subs{2}O, and \textit{Q}(HNC)/\textit{Q}(HCN) = 0.094 $\pm$ 0.023, about twice as high as the value of 0.05 predicted by the heliocentric dependence of \textit{Q}(HNC)/\textit{Q}(HCN) in comets found by \cite{Lis2008}.

\section{Analysis of Continuum Emission in ER61}\label{sec:cont}
\begin{figure}[ht!]
\plotone{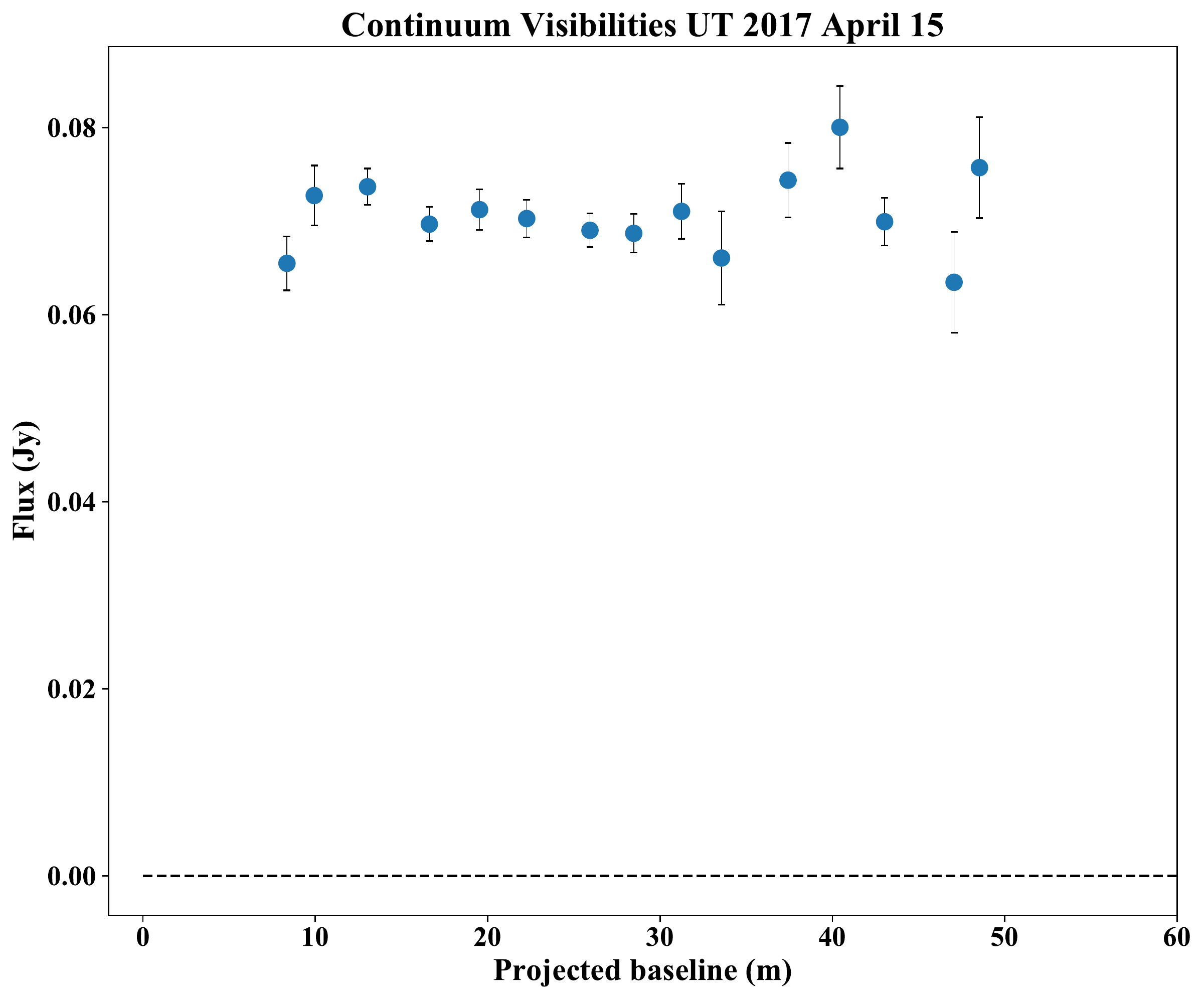}
\caption{Visibility amplitude vs. projected baseline for continuum  in ER61 on UT 2017 April 15. \label{fig:contvis}}
\end{figure}

The continuum detected on April 11 and 15 is compact and characteristic of unresolved thermal emission (Figures~\ref{fig:maps1} and~\ref{fig:maps2}) from a source much smaller than the interferometric beam ($\sim$ 2100 $\times$ 4700 km). Unlike the molecular line emission (Figures~\ref{fig:visplots1} and~\ref{fig:visplots2}) and as expected for an unresolved source, the visibilities for the continuum remain constant over the range of projected baseline lengths (9--49 m; Figure~\ref{fig:contvis}), with mean values $S_\lambda$ = 37 $\pm$ 5 mJy and 70 $\pm$ 4 mJy for April 11 and 15, respectively. The detected 0.85 mm emission is likely due to the thermal emission of a compact cloud of dust particles in the near-nucleus coma. Indeed, based on the Near-Earth Asteroid Thermal Model \citep[NEATM;][]{Harris1998} with a beaming factor $\eta$ = 0.7 and a bolometric emissivity of 0.8 \citep{Boissier2011}, the detected flux density of 37 mJy at 350 GHz on April 11 would imply an excessively large 19 km radius nucleus, inconsistent with other estimates of the nucleus size. Pre-discovery observations on 2014 February 12, when the comet was at 11.33 au from the Sun, place an upper limit on the nucleus radius of 7--10 km \citep{Meech2017}. In addition, both Pan-STARRS1 and WISE photometric observations performed in 2015, when the comet was at distances $\geq$ 6 au and weakly active, are consistent with a 10 km radius of the nucleus \citep{Meech2017}. The nucleus would then contribute at the level of 10.5 mJy to the continuum emission observed on April 11 (11.3 mJy on April 15). 

Comet ER61 underwent a 2 mag outburst that started on April 3.9 and peaked near April 6.6 UT \citep{Sekanina2017}. The ER61 ACA observations were thus obtained 7.7--11.5 days after the onset of this major outburst. Taking into account the elapsed time  between  the  outburst  onset  time  and  the ACA  observations, the compact coma observed at 0.85 mm could be explained if the dust particles mainly contributing to the observed emission had velocities significantly less than 4 m\,s$^{-1}$. This population of slowly moving dusty material would correspond to large debris resulting from  nucleus fragmentation, which was inefficiently accelerated by the gas pressure. The fast-moving, i.e. small particles, were observed in the optical. \cite{Opitom2019} collected dust brightness maps of comet ER61 at 710 nm on April 9 showing extended dust emission with $\sim$ 30 arcsec radius, that would correspond to projected dust velocities of $\sim$ 40 m\,s$^{-1}$. Maps of the dust color show a color gradient with the dust becoming bluer when further from the nucleus, consistent  with the natural dust size sorting expected after an outburst, that is, smaller particles dominating the outer coma and larger particles the inner coma \citep{Opitom2019}.    

The exceptional outburst of comet 17P/Holmes in 2007 October at \textit{r}\subs{H} = 2.4 au was observed with the Plateau de Bure interferometer on two dates, 3.8 and 4.9 days after the outburst onset \citep{Boissier2012}. The 3~mm dust continuum emission was resolved by the $\sim$ 7000 km projected synthesized beam, but was more compact and 20\% less intense on the latter date. The 17P data are best explained by the presence of two components: a slow-moving (7--9 m\,s$^{-1}$ for 1 mm particles), unresolved `core' component, and a high-velocity component (50--100 m\,s$^{-1}$ for 1 mm particles) with a steep size distribution corresponding to the fast expanding shell observed in optical images \citep{Boissier2012}. The compact dust cloud observed in comet ER61 seems to be analogous to the core component of comet 17P, except that its thermal emission increased by a factor two between April 11 and 15, whereas it remained approximately constant in 17P. The non-detection of the fast expanding shell in comet ER61 can be explained by its residence time within the ACA beam, of about a day for particles at 50 m\,s$^{-1}$.
     
The measured flux densities $S_\lambda$ allow us to derive the mass of the core component around ER61's nucleus, following:

\begin{equation}
S_\lambda = \frac{2 k T_{\rm dust} M_{\rm core} k_\lambda}{\lambda^2 \Delta^2}
\end{equation}
   
\noindent
where $T_{\rm dust}$ is the dust temperature (taken equal to the equilibrium temperature at the comet heliocentric distance, i.e. 259 K), and  
$k_\lambda$ is the dust opacity at the wavelength $\lambda$. The dust opacity depends on the dust size distribution and material optical properties. We used the opacity values computed by \cite{Boissier2012} at $\lambda$ = 0.8 mm for a size index of --3, a maximum size of 10 cm, and a dust porosity of 50\%. For optical constants corresponding to astronomical (amorphous) silicates or silicate mixtures made of 20\% amorphous silicates and 80\% forsterite, mixed with water ice in various proportions, the derived opacities are in the range 2--4 $\times$ 10$^{-2}$ m$^2$ kg$^{-1}$. Taking the mean value, we derive dust masses of $\sim$ 5 $\times$ 10$^{10}$ kg  and  8 $\times$ 10$^{10}$ kg for April 11 and 15, respectively. We did not consider the 15--30\% possible nucleus contribution to the measured $S_\lambda$. The dust masses are five times smaller if we set the maximum size to 1 cm. For comparison, using the same assumptions about the dust properties, the core of 17P Holmes is found to be 3--5 times more massive. The derived dust mass is a lower limit to the total mass released during the outburst of ER61 and does not include that of the fragment which was observed near the comet several weeks after the outburst  \citep{Sekanina2017}. 

\cite{Meech2017} determined a nucleus radius \textit{r}\subs{N} = 10 km, so our calculated dust mass on April 11 would constitute 0.0024\% of the nucleus mass. This corresponds to a fragment with a 288 m radius and suggests that a major fragmentation event occurred. The outburst duration (defined as the temporal distance from the onset to the peak brightness) is 2.6 days \citep{Sekanina2017}; therefore, the production rate related to the outburst is 5 $\times$ 10$^{10}$ kg / 2.6 days = 2.23 $\times$ 10$^5$ kg\,s$^{-1}$. There is no constraint on the dust production rate for the quiescent activity of ER61 in the literature. Given that the water production rate was $\sim$3587 kg\,s$^{-1}$ and assuming a dust/gas ratio of unity, the dust release rate was 62 times higher (average over 2.6 days) than during quiescent activity.

The factor of two increase of the flux density between April 11 and 15 (actually 1.9 $\pm$ 0.3 when not considering the nucleus contribution to the flux density, 2.2 $\pm$ 0.4 when the nucleus contribution is considered) is puzzling. A possible explanation is that the core is optically thick. In this case, we would expect its brightness to vary linearly with the projected surface area of the core and with the square of the elapsed time since the outburst onset, assuming that its size varies linearly with time (case of spherical expansion at constant velocity). Under this assumption, a factor of 2.2 increase is expected, which is consistent with the data. However, if the core is optically thick, its physical size can then be estimated for the two dates and its expansion velocity can be derived. Applying the NEATM model described above, we derive a 19 km radius for April 11, and a 25 km radius for April 15. The assumption of an optically thick coma yields a coma expansion velocity of only 0.02 m\,s$^{-1}$, much below the escape velocity at the nucleus surface of about 5 m\,s$^{-1}$ for a 10 km radius nucleus. Hence, the dust coma observed with ACA was most likely optically thin. A possible explanation is then dust fragmentation, which increases the dust cross-section for the same mass of material. Besides the nucleus breakup, \cite{Sekanina2017} suggested based on analysis of the light curves that the ejecta underwent fragmentation processes. Indeed, the extreme variability of the ER61 companion's activity suggests that it progressively shedded debris until its full disintegration by the end of 2017 June \citep{Sekanina2017}. 

Dust fragmentation is a highly complex process, and providing a comprehensive model of dust fragmentation in ER61 is beyond the scope of this work. However, we can comment on the residence time of particles in the ACA beam in this instance. As earlier noted, based on the dust velocity measured in 17P/Holmes with the Plateau de Bure interferometer, the dust velocity is 7--9 m\,s$^{-1}$ for 1 mm particles. For a velocity varying with particle size as a$^{-0.5}$, this results in velocities of 2.5 m\,s$^{-1}$ for 1 cm particles and 0.8 m\,s$^{-1}$ for 10 cm particles, giving residence times of 2.17 days, 7 days, and 22 days for sizes of 1 mm, 1 cm, and 10 cm particle radii, respectively. Inversely, if we impose that the residence time should be less than 11.5 d (the time between outburst onset and April 15) so that the same large-size material is present on both dates, we find a constraint that the particles should have a velocity less than 1.5 m\,s$^{-1}$. \cite{Sekanina2017} estimated that the escape velocity of the detected companion was 0.9 m\,s$^{-1}$, consistent with our calculated values.

\section{Conclusion}\label{sec:conclusion}
The powerful 12 m ALMA array has provided a window into the abundances and spatial distributions of volatiles in the inner coma with exceptional detail. Previous work with the main 12 m array has discerned parent from product species with high angular resolution maps of inner coma volatiles, revealed new insights into the origins of cometary HNC, probed the thermal physics of the inner coma, and detected the first parent volatiles in an interstellar comet \citep{Cordiner2014,Cordiner2017a,Cordiner2017b,Cordiner2020,Bogelund2017}. However, the 12 m array is not always in the ideal configuration during a cometary apparition. 

Due to the predetermined configuration schedule in any given observing cycle, the array can be in its more extended configurations during peak comet activity, where it resolves out extended flux and is blind to large-scale structure in the coma. Indeed, previous calculations have shown that the ACA is more sensitive to cometary flux than the 12 m array for configurations more extended than C43-4. However, this depends on the species lifetime, heliocentric and geocentric distance, as well as \lp{} (in the case of distributed sources). Still, the ACA is competitive with the 12 m array even in its most compact (C43-1) configuration. As an example, we consider a comet with \textit{Q}(H$_2$O) = 1 $\times$ 10$^{29}$ s$^{-1}$ at \rh{} = $\Delta$ = 1 au, with \textit{T}\subs{kin} = 70 K and a typical line width of 2 km\,s$^{-1}$. Using the ALMA Sensitivity Calculator, we find that in 3 hr of observing time the HCN (\textit{J} = 4 -- 3) transition could be securely detected (5$\sigma$) down to abundances of HCN/H$_2$O = 0.011\% for the ACA, HCN/H$_2$O = 0.003\% for the 12 m array in configuration C43-1, HCN/H$_2$O = 0.012\% for C43-4, and HCN/H$_2$O = 0.024\% for C43-5. Although these values are all consistent with strongly depleted abundances among measured comets, this highlights the increased sensitivity (by a factor of two for C43-5) of the ACA  for studying coma chemistry compared to more extended 12 m configurations.

The work presented here demonstrates the power of the 7 m ACA for cometary observations, providing a valuable tool for probing larger scales in the coma, highly complementary to the 12 m array. Our measurements indicate that HCN production was predominantly from nucleus sources in ER61, whereas CS, H\subs{2}CO, and HNC were produced by distributed coma sources. CH$_3$OH may have also been associated with a distributed source, perhaps icy grains, although our measurements cannot rule out direct nucleus release. Parent scale lengths on the order of 2200 km for H\subs{2}CO and 3300 km for HNC are consistent with previous work, suggesting that both molecules are produced from photolysis or thermal degradation of polymers, dust grains, and/or the organic macromolecular material found by Rosetta at comet 67P/Churyumov--Gerasimenko \citep{Capaccioni2015}. A parent scale length of 2000 km for CS is consistent with distributed source production of CS in the extended coma from an unknown parent and cannot be fully explained by CS$_2$ photolysis in the inner coma. Continuum emission in ER61 is consistent with unresolved thermal emission from slowly moving outburst ejectas. Assuming a maximum dust size of 10 cm, we derive dust masses of $\sim$ 5 $\times$ $10^{10}$ kg and 8 $\times$ $10^{10}$ kg for April 11 and 15, respectively. The factor of two increase in flux density between April 11 and 15 may have been due to dust fragmentation in the optically thin coma.

Our observations between 7.7 -- 11.5 days after ER61’s April outburst indicate that it is depleted in HCN and CO but consistent with the average for CS, HNC, CH$_3$OH, and H$_2$CO compared to abundances measured in its dynamical class. Our results show the wealth of information regarding molecular abundances, spatial distributions, mechanisms of molecular production, and coma temperatures that can be revealed in moderately bright comets such as ER61 with standalone ACA observations and highlight the important role that the ACA will play in future cometary studies.

\acknowledgments
This work was supported by the NASA Postdoctoral Program at the NASA Goddard Space Flight Center, administered by Universities Space Research Association under contract with NASA (N.X.R.), as well as the National Science Foundation (under Grant No. AST-1614471; M.A.C.), and by the Planetary Science Division Internal Scientist Funding Program through the Fundamental Laboratory Research (FLaRe) work package (S.N.M., M.A.C., S.B.C.), as well as the NASA Astrobiology Institute through the Goddard Center for Astrobiology (proposal 13-13NAI7-0032; S.N.M., M.A.C., S.B.C.). Part of this research was carried out at the Jet Propulsion Laboratory, California Institute of Technology, under a contract with the National Aeronautics and Space Administration. It makes use of the following ALMA data: ADS/JAO.ALMA \#2016.1.00938.T. ALMA is a partnership of ESO (representing its member states), NSF (USA), and NINS (Japan), together with NRC (Canada), MOST and ASIAA (Taiwan), and KASI (Republic of Korea) in cooperation with the Republic of Chile. The Joint ALMA Observatory is operated by ESO, AUI/NRAO, and NAOJ. The National Radio Astronomy Observatory is a facility of the National Science Foundation operated under cooperative agreement by Associated Universities, Inc. We thank an anonymous referee for feedback that improved the manuscript.

\appendix
\section{Effective Solar Pumping Rates}\label{sec:pumping}
The ACA measurements presented here sampled emission from nucleocentric distances of $\sim$2000--24,000 km (the angular resolution of the longest ACA baseline to that of the ACA autocorrelation primary beam). As suggested by the divergence of the best-fit rotational and kinetic temperatures, non-LTE effects are important and must be accounted for at such large nucleocentric distances. Although rotational states in the inner coma (several hundred km from the nucleus) are thermalized by collisions, rotational populations at larger nucleocentric distances are governed by fluorescence equilibrium. 

For a more complete treatment of the extended non-LTE coma sampled by our observations, fluorescence pumping by solar radiation must be considered, in which excited vibrational states pumped up by solar radiation decay through fluorescence, in turn further pumping the rotational levels in the ground vibrational state. We calculated effective solar pumping rates for CO, CS, CH$_3$OH, H$_2$CO, and HNC as in \cite{Zakharov2007} and \cite{Crovisier1983}, and scaled them by the heliocentric distance of ER61 at the time of our observations. HCN pumping rates were adopted from \cite{Cordiner2019}. We considered pumping by resonant fluorescence transitions (i.e., hot bands were excluded) for each species. We calculated pumping rates and Einstein-A's using the HITRAN molecular database \citep{Gordon2017} for CO, CS, CH$_3$OH, and H$_2$CO, and the GEISA molecular database \citep{Jacquinet-Husson2016} for HNC. Rotational levels were mapped from HITRAN or GEISA to the LAMDA molecular database \citep{Schoier2005} and the pumping rate for each level applied to our non-LTE radiative transfer models.


\bibliography{ER61_ACA}{}
\bibliographystyle{aasjournal}



\end{document}